\documentclass[epsf,prb,twocolumn,showpacs,nofootinbib,superscriptaddress]{revtex4-1}

\usepackage[pdftex]{graphicx}
\usepackage{dcolumn}
\usepackage{bm}
\usepackage{epsfig}
\usepackage{latexsym}
\usepackage{amsmath}
\usepackage{amsfonts}
\usepackage{amssymb}
\usepackage{color}
\usepackage{array}
\usepackage{framed}
\usepackage{braket}
\usepackage{comment}
\usepackage[normalem]{ulem}
\newcommand{\Mod}[1]{\ (\mathrm{mod}\ #1)}

\graphicspath{ {./figures/} }

\begin{document}

\title{Unitary-projective entanglement dynamics}

\author{Amos Chan}
\affiliation{Theoretical Physics, Oxford University, 1 Keble Road, Oxford OX1 3NP, United Kingdom}
\author{Rahul M. Nandkishore}
\affiliation{Department of Physics and Center for Theory of Quantum Matter,\\ University of Colorado, Boulder, CO 80309}
\author{Michael Pretko}
\affiliation{Department of Physics and Center for Theory of Quantum Matter,\\ University of Colorado, Boulder, CO 80309}
\author{Graeme Smith}
\affiliation{Department of Physics and Center for Theory of Quantum Matter,\\ University of Colorado, Boulder, CO 80309}
\affiliation{JILA, University of Colorado/NIST, Boulder, CO 80309}

\date{\today}

\begin{abstract} 
Starting from a state of low quantum entanglement, local unitary time evolution increases the entanglement of a quantum many-body system.  In contrast, local projective measurements disentangle degrees of freedom and decrease entanglement.  We study the interplay of these competing tendencies by considering time evolution combining both unitary and projective dynamics.  We begin by constructing a toy model of Bell pair dynamics which demonstrates that measurements can keep a system in a state of low ($i.e.$ area law) entanglement, in contrast with the volume law entanglement produced by generic pure unitary time evolution.  While the simplest Bell pair model has area law entanglement for any measurement rate, as seen in certain non-interacting systems, we show that more generic models of entanglement can feature an area-to-volume law transition at a critical value of the measurement rate, in agreement with recent numerical investigations.  As a concrete example of these ideas, we analytically investigate Clifford evolution in qubit systems which can exhibit an entanglement transition. We are able to identify stabilizer size distributions  characterizing the area law, volume law and critical `fixed points.' We also discuss Floquet random circuits, where the answers depend on the order of limits - one order of limits yields area law entanglement for any non-zero measurement rate, whereas a different order of limits allows for an area law - volume law transition.  Finally, we provide a rigorous argument that a system subjected to projective measurements can only exhibit a volume law entanglement entropy if it also features a subleading correction term, which provides a universal signature of projective dynamics in the high-entanglement phase.

Note: The results presented here supersede those of all previous versions of this manuscript, which contained some erroneous claims.
\end{abstract}

\maketitle

\normalsize

\tableofcontents

\section{Introduction}

The notion of quantum entanglement is a unifying theme across numerous areas of modern physics, from the study of solid state systems to the study of black holes.  In a condensed matter context, entanglement not only provides a window into the study of quantum ground states, but also is an important tool in characterizing the approach to thermal equilibrium (or the lack thereof).  For example, the entanglement entropy of a ground state or a many-body localized state usually obeys an area law, $S\sim \ell^{d-1}$, where $\ell$ is the linear size of the partition.  In contrast, the entanglement entropy of a generic thermalizing state at a non-zero temperature is given by a volume law, $S\sim \ell^d$.

While such static entanglement signatures are useful, there is also a great deal of information contained in the dynamics of entanglement.  Consider preparing a system in a tensor product state, for example by performing a quantum quench.  If the system exhibits many-body localization, then the growth of entanglement will be logarithmic in time, $S\sim\log t$.  In contrast, a generic thermalizing system will feature an initially ballistic ($i.e.$ linear) entanglement growth, eventually approaching a volume law, as expected for a thermal system.  Recently, studies have taken place on the growth of quantum entanglement under generic unitary time evolution, demonstrating in detail the linear growth of mean entanglement entropy, as well as determining the form of fluctuations around the mean.\cite{nahum1}  Subsequent analyses have studied both entanglement growth and spreading of local operators under random unitary time evolution, both with and without conservation laws.\cite{nahum2,khemani,curt1,curt2,fracop,banchi}  Similar work has also been done in the context of Floquet and Hamiltonian time evolution. \cite{amos1,amos2,amos3,jonay,nahum3,prosen1,prosen2,prosen3,sunderhauf}

But while unitary dynamics generically leads to the growth of entanglement, there is another more drastic type of time evolution which can decrease the entanglement of a quantum system.  Under certain conditions, such as interaction with a macroscopic classical object, a quantum mechanical system can rapidly evolve into an eigenstate of a specific operator, such that the resulting time evolution appears to be a non-unitary projection.  Such a process is referred to as a projective measurement.  When the system is projected into an eigenstate of a local operator, the corresponding local degree of freedom is disentangled from the rest of the system, resulting in a decrease in overall entanglement.  In this way, projective measurements can remove some of the entanglement created by more generic unitary time evolution.

Since unitary time evolution and projective measurements have opposite effects on entanglement, it is natural to ask how a physical system behaves when both types of evolution play a prominent role.  For example, a system could be subjected to a continuous series of measurements, as can be accomplished with superconducting qubits.\cite{qubit1,qubit2,qubit3}  As another potential physical realization, it has been proposed by M. Fisher that Posner molecules may play a role in quantum information processing in the brain.\cite{fisher,posner,halpern}  As these molecules bind and unbind, they undergo joint unitary-projective dynamics, generating entanglement between different molecules. 

In such a system with joint unitary-projective evolution, it is not obvious how the presence of projective measurements modifies the behavior of a purely unitary system.  It seems clear that measurement should decrease the steady-state entanglement entropy of the system.  But by how much?  $A$ $priori$, one possibility is that measurements might simply decrease the coefficient of the resulting volume law entanglement entropy.  In contrast, recent numerical investigations have indicated that a sufficient amount of measurements can limit entanglement entropy to area law scaling, behavior which is normally associated with ground states or many-body localized eigenstates.  For example, simulations of free fermion systems subject to continuous monitoring has indicated that an arbitrarily low rate of measurement is sufficient to keep the system in a state of low entanglement.\cite{fermion}  A hydrodynamic explanation for this behavior was also advanced.

\begin{figure}[t!]
 \centering
 \includegraphics[scale=0.5]{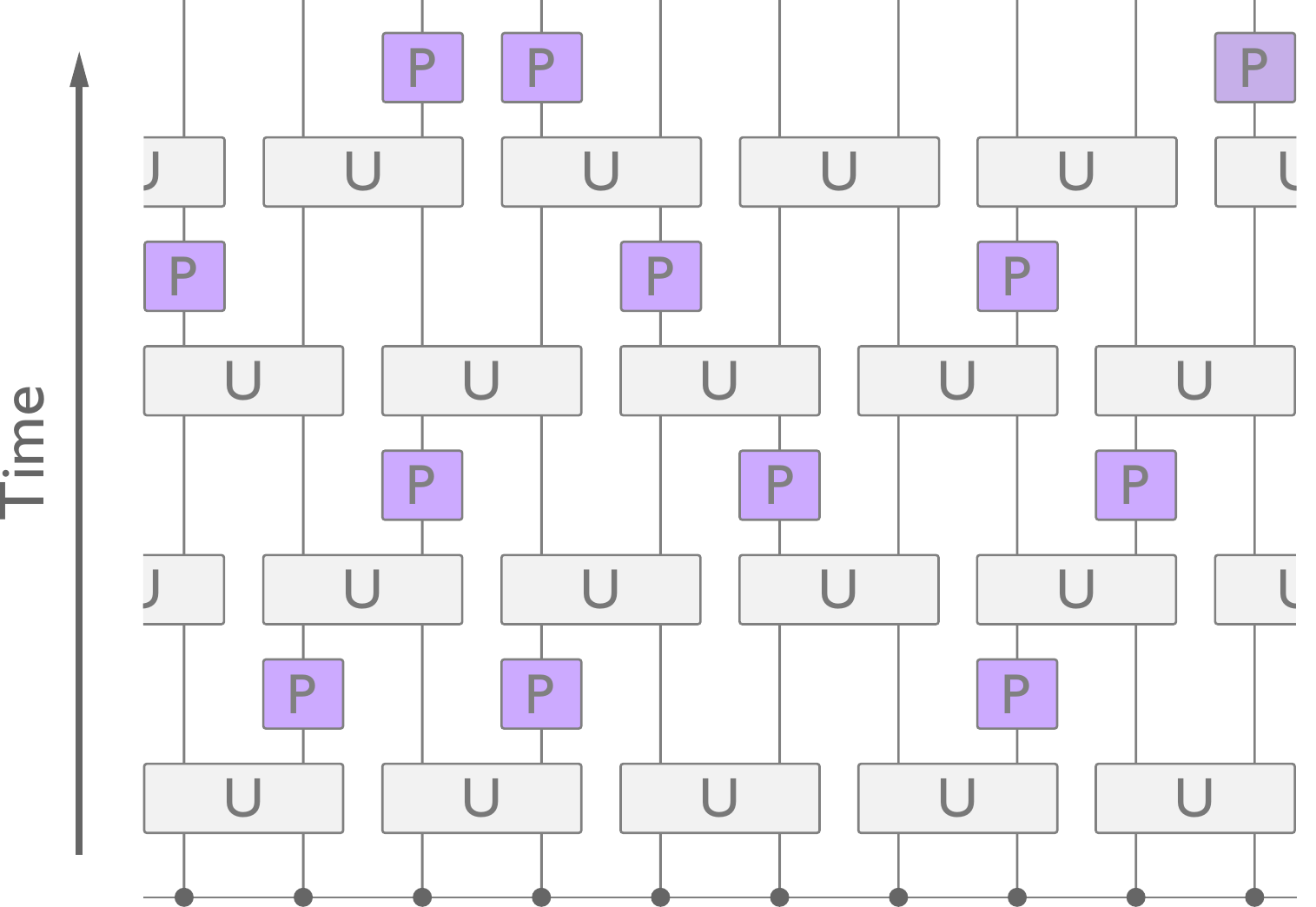}
 \caption{Unitary-projective time evolution can be modeled via a quantum circuit consisting of alternating layers of unitary and projective operators acting on a set of local degrees of freedom.  The grey rectangles represent local unitary operators which generate entanglement between adjacent degrees of freedom, while the purple squares represent local projective measurements which remove entanglement from the system.}
 \label{fig:circuit}
\end{figure}

More recently, multiple groups have investigated more generic unitary-projective time evolution, implemented in the language of quantum circuits.\cite{LiChenFisher,LiChenFisher2,SkinnerRuhmanNahum}  Specifically, one can consider time evolution via a circuit consisting of alternating layers of unitary and projective operators, as depicted in Figure \ref{fig:circuit}.  If every site were measured during each projective step, then the system would be continually reset to a tensor product state and no entanglement would ever be built up.  The more interesting scenario is when the local measurements are sparse.  Consider projective evolution in which each site has probability $f$ to be measured at each time step.  Equivalently, a randomly distributed fraction $f$ of the sites are measured at every step. As $f\rightarrow 1$, with every site being constantly measured, it is clear that the effects of projection dominate those of the unitary time evolution, and no significant entanglement should build up.  But for $f \ll 1$, there are far more unitary than projective operators in the circuit, indicating that the unitary evolution should proceed largely unaffected by projection, driving the system towards volume law entanglement entropy.  Consistent with this expectation, two independent groups have confirmed that such a model exhibits a transition from a high-entanglement volume-law phase at low $f$ to a low-entanglement area-law phase at $f$ close to $1$.\cite{LiChenFisher,LiChenFisher2,SkinnerRuhmanNahum}

In this work, we analytically investigate several models in which the effects of measurements on entanglement can be explicitly studied.  We begin in Section \ref{sec:toy} by constructing toy models for entanglement dynamics under unitary-projective evolution.  We first consider a simple model of Bell pair dynamics which illustrates how measurements can limit entanglement entropy to area law scaling.  The model is simple enough to allow an exact calculation of the dynamics of entanglement entropy.  Nevertheless, it captures a crucial piece of physics which determines the interplay between the two types of dynamics: local unitary evolution primarily creates short-range entanglement, while projective measurements can destroy entanglement on any length scale.  The result is that the system is dominated by small Bell pairs, leading to an area law for entanglement entropy.  By continually removing long-range entanglement from the system, projective measurements are able to keep the entire system in a state of unusually low entanglement.  We also investigate the dynamics of entanglement, finding an overshoot phenomenon, whereby at intermediate times entanglement entropy exceeds its steady state value.

For the simple Bell pair model, area law behavior holds for any finite measurement rate $f$, as seen in the context of free fermion systems.  However, we show that modifications to this model to include more realistic entanglement patterns results in an area-to-volume phase transition at a finite critical measurement rate, consistent with numerics on unitary-projective quantum circuit evolution.  Specifically, we construct a model in which clusters of spins are all mutually entangled with each other, instead of the simple two-body entanglement of the Bell pair model.  We determine a differential equation governing the size of such clusters, which we find has exponentially decaying solutions (indicating area-law entanglement entropy) only for a critically large measurement rate.  For lower measurement rates, the size of such clusters will keep increasing until a volume-law entanglement entropy is reached.  This model thereby exhibits a concrete example of an area-to-volume law phase transition.

With these toy models in hand, we move on to test our intuition in more concrete situations with analytically tractable dynamics.  First, in Section \ref{sec:clifford}, we consider Clifford evolution in a qubit system, in which the unitary layers of the dynamics have random operators drawn only from the set of Clifford gates.  While this is not a universal set of gates, Clifford evolution allows for a convenient description of entanglement spreading in terms of an effective hydrodynamics.  We show that random Clifford evolution exhibits an area-to-volume law transition at a finite critical measurement rate.  This transition can be characterized in terms of the size of stabilizer generators, which are relatively small for high measurement rates, then jump to the size of the system at low measurement rates.  We derive a differential equation governing the stabilizer size distribution.  This differential equation predicts the existence of both area- and volume-law phases, as well as a critical point with a logarithmic area-law violation.  We also propose a hydrodynamic description for entanglement growth within the area-law phase.

In Section \ref{sec:floquet}, we investigate another type of analytically tractable model in the form of two Floquet random circuits with large on-site Hilbert space dimension. In these circuits, the Renyi-$\alpha$ entropies for $\alpha \geq 2$ can be mapped to emergent statistical mechanics problems, which amount to enumerating  minimal-length domain wall diagrams. The longer the lengths of the domain walls in these diagrams are, the higher the averaged entanglement entropy of these circuits will be.  In the regime at infinite $q$ and finite but arbitrarily large $L$, an area-law saturation of higher Renyi entropies results from the fact that projective measurements can provide effectively $L$ ``free'' segments of domain walls, along which no amount of entanglement entropy is associated.  This indicates an area-law phase for any finite measurement rate, consistent with the Bell pair model.  However, the conclusions are sensitive to the order of limits, and a different order of limits may allow for an area-to-volume law transition at a finite critical measurement rate.

In Section \ref{sec:General}, we present certain general arguments constraining the form of entanglement entropy in the presence of measurements.  We find that, while an area law entanglement entropy can exist without any special extra structure, a volume law entanglement entropy can only exist in the presence of a subleading correction.  
While the results in this section are all for Von Neumann entropy, Renyi entropies $S_n$ with $n>1$ are all upper bounded by Von Neumann entropy, so these results serve as useful upper bounds on all higher Renyi entropies.

Finally, in Section \ref{sec:conclusions}, we summarize our results and outline certain future directions of investigation opened by our work.

\emph{Note:} A previous version of this manuscript made the erroneous claim that area-law entanglement entropy was generically present for any nonzero measurement rate. This is true in some models (which constitute a particular universality class) but is not true for all models. This mistake has been corrected in the more detailed analysis of the present version, the results of which supersede all previous versions.

\section{Toy Models for Entanglement Dynamics}
\label{sec:toy}

\subsection{Bell Pair Model}

\begin{figure}[t!]
 \centering
 \includegraphics[scale=0.4]{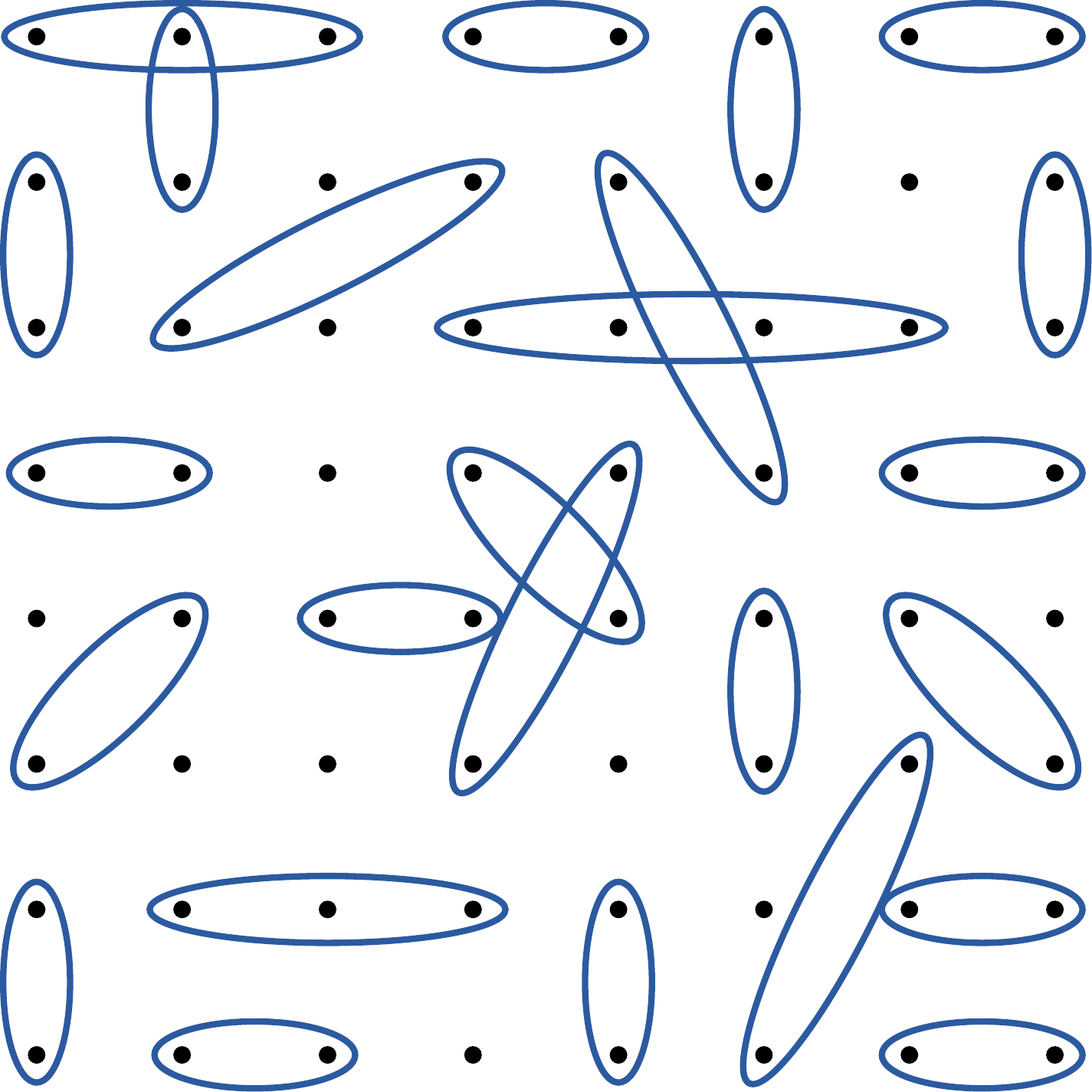}
 \caption{We study a qubit system in which each qubit either forms a maximally entangled Bell pair with another qubit (represented by an oval) or is completely disentangled from the rest of the system.}
 \label{fig:bell}
\end{figure}

In order to build intuition for unitary-projective dynamics, it is useful to construct a toy model which captures some of the important physical features.  To this end, we first focus on a particularly simple form of entanglement.  We consider states which can be fully described in terms of Bell pairs, $i.e.$ maximally entangled two qubit states, such as a spin singlet.  In other words, we study a system of qubits in which each qubit is either maximally entangled with another qubit or is completely disentangled from the system (see Figure \ref{fig:bell}).  For such a system, we can easily obtain the entanglement entropy by counting the number of Bell pairs which are cut by a given partition.  While such Bell pair configurations are a restricted class of states, this model will provide important intuition as to how measurements can restrict entanglement entropy to area law scaling.  All Renyi entropies are equal for this model. 

To build unitary-projective dynamics into the toy model, we must consider the effects of both types of operators on Bell pairs.  We first consider applying a layer of local unitary operators, as in Figure \ref{fig:circuit}.  Such a layer of operators will result in entanglement between neighboring qubits which were previously disentangled from the rest of the system.  Consistent with the restrictions of our toy model, we take this entanglement to be maximal.  In other words, local unitary operators can generate Bell pairs between previously unentangled neighboring qubits.  When a unitary operator acts on a qubit which was already in a Bell pair, it can move one end of the Bell pair to an adjacent site, which may cause the Bell pair to grow or shrink in size.  Bell pairs can move through one another. Starting from a state with mostly small Bell pairs ($i.e.$ a state of low entanglement), generic local unitary time evolution will cause Bell pairs to increase in size, leading to the growth of entanglement for generic spatial partitions.

While local unitary operations tend to increase entanglement, via creating small Bell pairs which subsequently grow in size, the projective portion of the time evolution has a radically different effect.  Performing a projective measurement on a qubit has the effect of disentangling it from the rest of the system.  If that qubit happened to be in a Bell pair with another qubit, that Bell pair is destroyed by the measurement.  Notably, this mechanism for Bell pair destruction is equally effective for Bell pairs of any size, since a local measurement on either qubit is sufficient to destroy a Bell pair, regardless of the distance to the other qubit.  The model thereby captures the expected interplay between unitary and projective dynamics: creation of short-range entanglement (and its subsequent growth) by unitary operators, coupled with removal of entanglement at all length scales by projective measurements.

Combining these two types of physical processes, we can now very easily write down a set of equations governing the time evolution of the distribution $P(x)$ on the spatial size $x$ of Bell pairs.  At each time step, the local unitaries cause a given Bell pair to either grow (with probability $p_g$), shrink (with probability $p_s$), or remain the same size (with probability $(1-p_g-p_s)$).  We also take a fraction $0<\tilde{f}<1$ of the Bell pairs to be destroyed, $i.e.$ be reset to zero size, where $\tilde{f} = 2f-f^2$ is the probability of the Bell pair being measured on at least one of its two sites.  Away from $x=0$, the time evolution equation for $P(x)$ takes the form:
\begin{align}
\begin{split}
\partial_t P(x) = -P(x)& + (1-\tilde{f})\bigg(p_s P(x+1) +\\ &p_g P(x-1) + (1-p_g-p_s)P(x)\bigg)
\end{split}
\end{align}
where the time step (defined by one layer of unitaries and one layer of projectors) is taken to be $1$.  The probabilities $p_g$ and $p_s$ will not depend on the rate of external measurement, and therefore have no $\tilde{f}$ dependence.  We also neglect any nonlinearities of this equation, such that the probabilities can be taken to be independent of $P(x)$.  The probabilities can, however, generically depend on the size $x$.  We can now take the continuum limit of the above time evolution equation to obtain:
\begin{equation}
\partial_t P =  - (1-\tilde{f})\gamma\partial_xP - \tilde{f}P
\label{master}
\end{equation}
where $\gamma = p_g-p_s$ is the difference in probabilities for growing and shrinking of a Bell pair.  The first term on the right, which is the only term present at $f=0$ ($i.e.$ pure unitary evolution), would lead to uni-directional propagation of waves in the distribution, with the direction of propagation depending on the sign of $\gamma$.  However, the second term on the right, arising from the projective measurements, causes the distribution to decay, preventing entanglement from propagating very far from $x=0$.  To have a steady state solution, with $\partial_tP(x) = 0$, the distribution must satisfy:
\begin{equation}
\partial_xP = -\frac{\tilde{f}}{\gamma(1-\tilde{f})}P
\end{equation}
If we take $\gamma$ to be approximately independent of $x$, then we immediately obtain:
\begin{equation}
P(x)\sim e^{-\lambda x}
\label{asymp}
\end{equation}
where $\lambda = \tilde{f}/\gamma (1-\tilde{f})$.  Note that we have not needed to make use of the details of what happens to the distribution at $x=0$, $i.e.$ the details of Bell pair creation at small scales, which only serves to determine the behavior near the origin.

We see that the steady state solution of the joint unitary-projective time evolution is dominated by small Bell pairs, such that the system is mostly short-range entangled.  This makes intuitive sense, in that long-range entanglement is being constantly removed from the system by projective measurements, while the entanglement resulting from local unitary evolution is only being created on short scales.  We can directly calculate the typical entanglement entropy of the system.  For example, consider a one-dimensional system, which we partition into two half-lines.  A qubit at a distance $x$ from the cut will contribute one bit ($i.e.$ $\ln 2$) to the entanglement entropy if it is a member of a Bell pair of size at least $x$, and if that Bell pair extends in the direction of the cut.  Summing contributions from all qubits on one side of the partition, we obtain:
\begin{equation}
S\sim \int_0^\infty dx\int_x^\infty dx'P(x')\sim\textrm{constant}
\end{equation}
The entanglement entropy is a constant, $i.e.$ independent of the system size $L$, since the exponentially decaying distribution $P(x)$ yields a convergent integral.  Since the entanglement entropy is constant, we conclude that the asymptotic state of the unitary-projective evolution obeys an area law, as opposed to the $S\sim L$ behavior of a volume law state.  In higher dimensions, we will have the same sort of exponential convergence of the entropy integrals, except with a factor of area arising from integrating over the entire partition.  In this way, our Bell pair toy model gives rise to an area law for entanglement entropy in any dimension.

\begin{figure}[t!]
 \centering
 \includegraphics[scale=0.37]{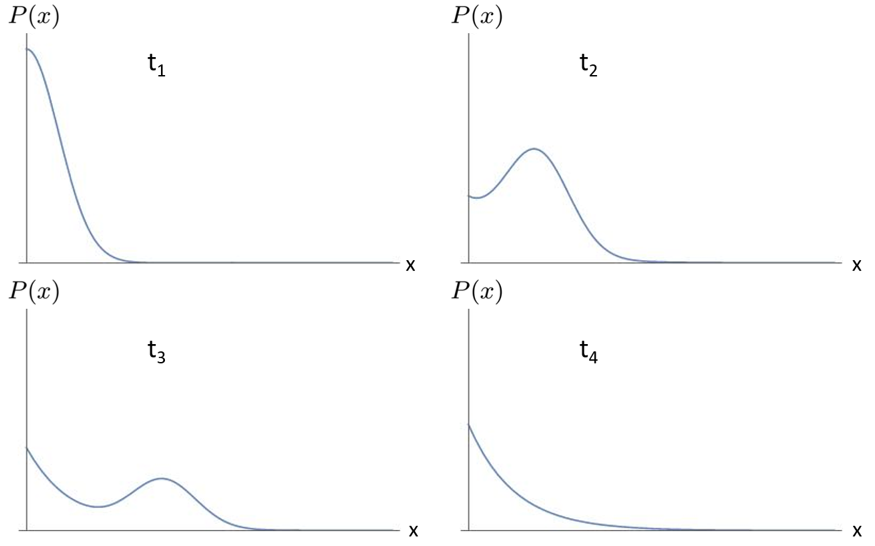}
 \caption{A schematic representation of $P(x)$ at various times for an initial tensor product state.  The evolution is characterized by a ballistically propagating but decaying peak, as well as the development of an exponential distribution near $x=0$.}
 \label{fig:peak}
\end{figure}

\begin{figure}[t!]
 \centering
 \includegraphics[scale=0.45]{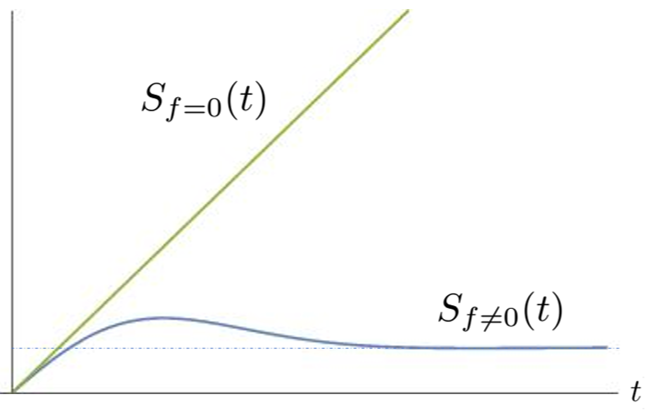}
 \caption{For a nonzero projection probability $f$, the entanglement entropy of an initial tensor product state asymptotes to an area law (after an initial overshoot), in contrast with the unitary ($f=0$) system, in which entanglement continues to grow towards a volume law state.  For $f\neq 0$, the initial growth is linear, but with a slope less than that of the pure unitary system.}
 \label{fig:entropy}
\end{figure}

In addition to the steady state, it is also easy to obtain the full time evolution of the Bell pair distribution.  The generic solution to Equation \ref{master} takes the form:
\begin{equation}
P(x,t) = e^{-\lambda x}g(x-vt)
\end{equation}
where $\lambda = \tilde{f}/\gamma(1-\tilde{f})$ and $v=\gamma(1-\tilde{f})$, while the function $g$ is an arbitrary function of $x-vt$.  (This form holds only away from $x=0$, near which the behavior will be modified in a complicated way in order to preserve the overall normalization of $P(x,t)$.  Note also that we should demand that $g$ grow no faster than exponentially, such that $P$ remains normalizable.)  The resulting time evolution takes the form of waves which propagate at velocity $v$, while decaying via the exponential factor $e^{-\lambda x}$.  For example, let us consider an initial tensor product state, such that all the weight of $P(x,t=0)$ is concentrated at $x=0$ and the entanglement entropy is zero.  As time evolves, the peak at $x=0$ propagates to the right at speed $v$, just as in the case of pure unitary evolution.  For short times ($t\ll 1/\lambda v$), the entanglement entropy will therefore grow linearly, $S\sim v_Et$, with an effective entanglement velocity given by:
\begin{equation}
v_E = \gamma (1-\tilde{f})
\end{equation}
We see that the initial entanglement velocity of this unitary-projective system is smaller than that of a pure unitary system by a factor of $(1-\tilde{f})$.  As time evolves, however, the slowdown of entanglement growth becomes more severe, as the weight in the propagating peak decays exponentially and is transferred back to the origin, as depicted in Figure \ref{fig:peak}.  (In a more generic dynamical model, the peak would begin to broaden as time evolves, though this is unimportant for present purposes.)  The contribution to the entanglement entropy from the decaying ballistic peak behaves as $S_{\textrm{ballistic}}\sim te^{-\lambda v t}$, which has a maximum value around $t_{\textrm{max}}\sim 1/\lambda v$, after which the entanglement entropy decreases to its area law saturation value, set by the exponential distribution near the origin.  The schematic behavior of the entanglement entropy as a function of time is depicted in Figure \ref{fig:entropy}.

\subsection{Cluster Model}

In the previous subsection, we encountered a toy model for entanglement dynamics under unitary-projective time evolution which featured area-law entanglement entropy for any nonzero measurement rate.  However, the persistence of the area-law down to arbitrarily low measurement rates may be an artifact of the restricted nature of the Bell pair toy model, which captures only a limited set of entanglement patterns.  In this subsection, we generalize the previous model to account for slightly more general types of entanglement.  Specifically, we allow for larger clusters of mutually entangled spins, instead of the simple two-body entanglement of the Bell pair model.  We find that this generalized model features an area-to-volume law transition at a nonzero critical measurement rate.  Thus, while the Bell pair model provides a good description of the novel area-law phase, it is not sufficiently powerful to describe the details of the area-to-volume law phase transition.

To construct a more general model of entanglement than the simple Bell pair model, we allow each spin to be entangled with any number of other spins, as opposed to being entangled with only a single other spin.  However, starting from a product state, entanglement will still be built up locally.  At short times, a given spin will only be significantly entangled with some local cluster of spins.  We work in one dimension for simplicity, labeling the size of a cluster of entangled spins as $x$.  Note that a Bell pair of size $x$ is one special example of a cluster of size $x$, in which only the two end spins are entangled.  More generally, however, all spins within a given cluster will be entangled with each other.  As in the Bell pair model, we can describe the system by a probability distribution of the size of clusters, $P(x)$.

As with Bell pairs, local unitary evolution will tend to increase typical size of clusters, which grow in a local fashion, spreading ballistically.  Also, as before, measurements will tend to decrease the size of clusters.  However, whereas measurement on a Bell pair automatically decreases its size from $x$ to $0$, measurement on a terminal spin of an entangled cluster can drop the size of the cluster to anywhere between $0$ and $x$.  (Note that measurement on the interior of the cluster cannot decrease its size.)  For a measurement on a terminal spin to drop the size of a cluster from $x$ to $x'$, all spins between $x$ and $x'$ must be disentangled from the rest of the cluster.  The probability of having all spins between $x$ and $x'$ disentangled from the rest of the cluster decays rapidly, as an exponential function of $x-x'$.  As such, we can very simply modify the differential equation governing $P(x)$ as follows:
\begin{equation}
\partial_t P = -(1-\tilde{f})\gamma\partial_xP - \tilde{f}P + \tilde{f}\int_x^\infty dx' P(x')e^{-\mu (x'-x)}
\label{newdiff}
\end{equation}
for some parameter $\mu$, where the final term represents the growth of $P(x)$ due to clusters of size $x'$ dropping to size $x$.

With this new toy differential equation in hand, we consider its implications for entanglement.  Let us first assume, as before, that the steady state distribution is exponentially decaying, $P(x)\sim e^{-\lambda x}$.  Plugging in this ansatz, Equation \ref{newdiff} implies that we must have:
\begin{equation}
\lambda(1-\tilde{f})\gamma - \tilde{f} + \frac{\tilde{f}}{\lambda+\mu} = 0
\end{equation}
As $\mu\rightarrow\infty$, this reduces to the results of the previously studied Bell pair model, with $\lambda = \tilde{f}/\gamma(1-\tilde{f})$, which is real and positive for any value of $\tilde{f}$.  However, as $\mu$ decreases, this equation will eventually cease to have real solutions.  Through straightforward algebra, it can be checked that there are only real solutions for $\lambda$ when the measurement rate satisfies $\tilde{f}>\tilde{f}_c$, where the critical measurement rate is:
\begin{equation}
\tilde{f}_c = \frac{\gamma \mu^2}{2-2\sqrt{1-\mu}-\mu +\gamma\mu^2}
\end{equation}
For $\tilde{f}>\tilde{f}_c$, we will therefore have only short-range entanglement, resulting in an area law for entanglement entropy.  For $\tilde{f}<\tilde{f}_c$, however, an exponential cannot be a steady state solution, but rather will feature a growing entanglement.  Similarly, it is easy to show that no power-law decay can provide a steady state solution when $\tilde{f}>\tilde{f}_c$.  In this regime, therefore, $P(x)$ will run off towards being close to uniform, thereby resulting in a maximally entangled volume law state.  This toy model thus demonstrates that area law behavior need not persist down to arbitrarily low measurement rates, as in the Bell pair model.  More generically, a system may exhibit an area-to-volume law transition at a finite critical measurement rate.

\section{Clifford Evolution in One Dimension}
\label{sec:clifford}

While the toy models in the previous section are useful for building intuition, it is important to have more concrete models with analytically tractable dynamics in which we can observe the same physics.  We now test our intuition in a model of a one-dimensional system evolving via an almost-random set of unitary gates.  Specifically, we consider the unitary operators of the time evolution to be randomly drawn from the set of Clifford gates, a form of dynamics referred to as Clifford evolution.  While this is not a universal set of gates, Clifford evolution in one dimension provides certain convenient technical simplifications while still capturing most of the qualitative features of truly random unitary evolution.  It also has the virtue that all Renyi entropies behave the same way. 

Clifford evolution relies on a simple action of Clifford gates on states labeled in terms of stabilizers, $i.e.$ operators $O_i$ which leave the state invariant, such that $O_i|\psi\rangle \sim |\psi\rangle$.  For a one-dimensional system with $L$ sites, $L$ such stabilizers will be necessary to fully label the state.  For example, if $|\psi\rangle$ is a tensor product state, then all $O_i$ will be local operators, acting non-identically only on a single site.  For an entangled state, the stabilizer operators extend over multiple sites, with the size of stabilizers increasing as the state becomes more entangled.  The value of Clifford evolution lies in its simple action on Pauli operators, mapping each Pauli to a product of other Pauli operators.  If we begin from a tensor product state labeled by a Pauli stabilizer on each site, then the resulting time evolution is simply described in terms of $L$ Pauli strings. For a given stabilizer state, the entanglement across a cut between $A$ and $B$ is determined as follows.  The stabilizer group $S$ for a state on $AB$ can be generated by three subgroups: $S_A$ and $S_B$ consist of stabiilzer elements with support on only $ A$ or $B$, respectively, while $S_{AB}$ contains stabilzers acting on both $A$ and $B$ and accounts for correlations between the systems.  A set of generators for $S$ is called minimal if it contains the minimal number of generators acting on both $A$ and $B$, and the total number of such generators is $|S_{AB}|$ (the size of the minimal generating set of the nonlocal group $S_{AB}$).  The entanglement between $A$ and $B$ is then $|S_{AB}|/2$ (see Ref. \onlinecite{Fattal} for details).

\subsection{Stabilizer Size Distribution}

We now consider entanglement growth in a system subject to Clifford evolution, starting from a direct product state, such that all stabilizers are of size $1$.  Suppose a random Clifford circuit is run on this system for some time $O(w)$, so that the typical weight of a minimal stabilizer generator is $\approx w$.  We call its stabilizer group $S$.  The stabilizer generators of $S$ will have support that is spatially localized to a region of width $\sim w$, and their supports will be distributed uniformly in space.  What is the effect of a single measurement at site $i$?  Suppose we measure at a fixed site $i$. Let the probability of a weight $w$ stabilizer be $P_w$.  We want to understand the effect of measurements on the distribution of stabilizers weights.  The total number of stabilizers of weight $w$ is $n_w = L P_w$. Typically, these stabilizers are uniformly spread out in the system, and the number of stabilizers of weight $w$ intersecting site $i$ is the density of stabilizers of weight $w$, multiplied by the weight $w$, i.e. $n_w w /L = P_w w$. The total number of stabilizers of weight less than $w$ that intersect $i$ is $\sum_{w'=1}^w P_{w'} w'$.

Given a stabilizer whose weight intersects site $i$, the probability that such stabilizer commutes with $Z_i$ is around $1/2$.  Therefore, the probability that $w$ is the lowest size of stabilizers that anti-commute with the measurement Pauli operator $Z_i$ is:
\begin{equation}
\mathrm{Prob}_\mathrm{lowest} (w) = \frac{1}{2^{\sum_{w'=1}^w P_w' w'}} 
\label{decprob}
\end{equation}
The measurement update amounts to throwing out the stabilizer with the lowest possible size which anticommutes with $Z_i$, and multiplying all the other stabilizers commuting with $Z_i$ by the stabilizer we throw away.

Now we want to determine the steady-state distribution of stabilizers subject to unitary-projective evolution.  Accounting for the weight-dependent probability of a nontrivial stabilizer being removed by a measurement, we can easily write down a differential equation describing the time evolution of $P_w$, just as in the Bell pair model.  In the continuum limit, the evolution of $P_w(t) $ will be described by:
\begin{equation}
\partial_t P_w(t) = - ( 1- \tilde{f}) \gamma \partial_w P_w(t) - \tilde{f} P_w(t)
\label{clifdif}
\end{equation}
where the effective measurement rate $\tilde{f}$ is given by:
\begin{equation}
\tilde{f} = \frac{f}{\exp[ \int_{0}^{w} P_{w'} (t) w' dw']}
\label{feff}
\end{equation}

The steady state solution requires:
\begin{equation} \label{Eq:steady}
\partial_w P^{(\mathrm{s})}_w =  \frac{ - \tilde{f} P^{(\mathrm{s})}_w }{ (1-\tilde{f})\gamma}
\end{equation}
One simple ansatz for the large $w$ behavior is an exponential distribution $P(w) = \lambda e^{ - \lambda w}$ which leads to an area law.  Plugging this into Equation \ref{feff}, we find that the effective measurement rate is given by:
\begin{equation}
\tilde{f} = \frac{f}{\exp(1/\lambda)}
\end{equation}
where we have dropped a term in the denominator which is exponentially small at large $w$.  Plugging this effective rate into Eq.~\ref{Eq:steady} gives the condition:
\begin{equation} 
\lambda = \frac{fe^{-1/\lambda}}{\gamma (1-fe^{-1/\lambda})}
\end{equation}
This equation only has solutions for large $f$, with the critical value $f_{c}$ given by:
\begin{equation}
f_c = \frac{\lambda^2e^{1/\lambda}}{\lambda^2 - \lambda + \gamma^{-1}}
\end{equation}

At smaller $f$, there is no exponentially decaying solution to the exponential. Instead, let us examine the large-$w$ ansatz $P^{(\mathrm{s})}_w = C w^{-2}$. Plugging this form for $P_w$ into Eq.~\ref{Eq:steady}, we have:
\begin{equation}
-2 C w^{-3 } = \frac{ -\gamma ' C w^{-2}}{ \exp (C \log(w/a) + a^2) } 
= -C e^{-a^2 } \gamma' a^{C} w^{-(2+C)}
\end{equation}  
where $a$ is a short-distance cutoff representing the scale at  which power-law decay sets in.  Note that we have set $(1-\tilde{f})\approx 1$ at large $w$, since the effective measurement rate now decays as a function of $w$.  It can be easily checked that this form for $P_w$ is a solution to the differential equation when $C=1$ and $\gamma' a = 2$.  For this solution, the entanglement entropy behaves as 
\begin{equation}
S = \int_{0}^{\infty} dw  \int_{w}^{\infty } dw' P^{(s)}_{w'} \sim \log L
\end{equation}
which is a log violation of the area law.  However, it is easy to see that this log violation represents only a critical point of the Clifford dynamics.  If we had chosen $P_w\sim w^{-n}$ for $n>2$, then it is easy to check that the Equation \ref{clifdif} implies that the distribution decays until it reaches a short-ranged exponential, corresponding to area-law entanglement entropy.  In contrast, if we had tried $P_w\sim w^{-n}$ for $n<2$, then the differential equation implies that the distribution will keep increasing until it is almost uniform, corresponding to volume-law entanglement entropy.  We therefore see that Clifford dynamics in one dimension possesses two fixed points, corresponding to area-law and volume-law phases, as well as a transition between them at a critical measurement rate $f_c$.  These results are consistent with both our cluster model and also recent numerical work on Clifford circuits. Finally, a distribution of the form $P(\omega) \sim \frac1L[1+B\exp(-\omega^2/L)]$ provides an asymptotic `high entanglement' solution to the steady state equation in the regime $\sqrt{L} < \omega < L$, and provides a partial characterization of the `volume law' phase. Note that this form of distribution generically produces an entropy equal to an volume law with an additive logarithmic correction. 

\subsection{Quasiparticle Picture and Hydrodynamics}

The effects of measurement in the context of Clifford evolution can also be understood using a representation in terms of a set of fictitious ``particles," as developed in Reference \onlinecite{nahum1}, which also allows a slightly more refined analysis.   We briefly recap the central idea behind the particle representation of Clifford evolution, referring the reader to Reference \onlinecite{nahum1} for further details.

\begin{figure}[t!]
	\centering
	\includegraphics[width=0.47\textwidth]{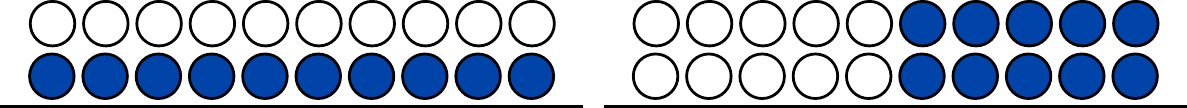}
	\caption{Clifford evolution can be represented in terms of a set of fictitious particles (blue circles) representing the endpoints of stabilizers.  The system contains as many particles as sites, and no more than two particles can occupy any site.  A tensor product state corresponds to a state with uniform density, while a maximally entangled state corresponds to all particles shifted to one side.  Figure adapted from Reference \onlinecite{nahum1}.}
	\label{fig:balls}
\end{figure}

For many purposes, it is sufficient to only keep track of the endpoints of the stabilizer, which encode information about the length of the Pauli strings.  To this level of detail, we can represent a state by a set of fictitious ``particles" representing the stabilizer endpoints, as depicted in Figure \ref{fig:balls}, where blue circles represent right endpoints and white circles represent left endpoints.  It can be shown that, due to a gauge freedom in choosing the stabilizers labeling the state, the total number of endpoints (left plus right) on a site can be chosen to be exactly two.  It is then sufficient to only keep track of the right endpoints (blue circles), while the left endpoints can be regarded simply as ``holes."  Within this representation, a tensor product state corresponds to a uniform density of particles, since each site is the left and right endpoint of a local (on-site) stabilizer.  Entanglement is then represented as a deviation from this uniform density.  Indeed, as discussed in Reference \onlinecite{nahum1}, the entanglement entropy associated with a partition at location $x$ is given by:
\begin{equation}
S(x) = \sum_{i>x}(\rho_i - 1)
\end{equation}
In other words, the entanglement entropy is given by the excess particle number on one side of the partition.  As the system evolves under random unitary time evolution, the tendency is for stabilizers to grow, which amounts to particles ($i.e.$ right endpoints of stabilizers) to drift to the right.  As shown in Reference \onlinecite{nahum1}, unitary evolution causes the particles to undergo biased diffusion, such that the density evolves as:
\begin{equation}
\partial_t\rho = \nu\partial_x^2\rho + \frac{\Lambda}{2}\partial_x((\rho-1)^2) - \partial_x\eta
\end{equation}
where $\nu$ and $\lambda$ are constants, and $\eta$ is a random variable representing noise.  Eventually, pure unitary evolution would take the system to a maximally entangled state, in which all particles are as far to the right as possible, as seen in Figure \ref{fig:balls}.

However, this flow of particles to the right is interrupted in the presence of projective dynamics.  The effect of local measurements is to disentangle spins from the rest of the system, which corresponds to the destruction of stabilizers ($i.e.$ resetting them to length $1$).  We can easily account for the removal of nontrivial stabilizers in the diffusion equation by adding a decay term on deviations from the mean density:
\begin{equation}
\partial_t\rho' = \nu\partial_x^2\rho' + \Lambda\rho'\partial_x\rho' - \tilde \lambda_\rho \rho' - \partial_x\eta
\label{densevol}
\end{equation}
where $\rho' = \rho-1$.  Based upon the discussion from the previous subsection, the decay constant $\tilde{\lambda}_\rho$ should depend on the quasiparticle distribution $\rho$, which in turn is set by the stabilizer distribution.  However, let us focus our attention for now to the area-law phase, for which we assume that the stabilizer distribution is short-ranged, such that the exponential factor in Equation \ref{decprob} is negligible.  Then we can safely take $\tilde \lambda_\rho \sim f$ to be a constant.  The differential equation for $\rho'$ now takes the form of a non-linear diffusion equation, where the diffusing density can decay with constant probability. It is exponentially unlikely that significant density will diffuse very far to the right of the cut, as expected for the area law phase.

We now propose a hydrodynamic description for entanglement under unitary-projective dynamics within the low-entanglement area-law phase.  We begin with the observation from Reference \onlinecite{nahum1} that under random unitary time evolution, the entropy on a given site $S(x)$ evolves according to\begin{equation}
\partial_t S = D\partial_x^2S + 1-(\partial_xS)^2 + \eta
\label{evolve}
\end{equation}
where $\eta$ is a noise term.  We now wish to also account for the effects of projective measurements.  When a site is measured, it becomes disentangled with the rest of the system, such that there is no difference in entanglement between the partitions to the immediate left and right of that site.  In other words, measurement sets the value of $\partial_xS$ to zero on that site.  The projective portion of the evolution then acts as a decay term on the evolution of $\partial_xS$, which does not have a natural local form in terms of $S$.  We therefore take a derivative of Equation \ref{evolve} and add an appropriate decay term to yield:
\begin{equation}
\partial_tS' = D\partial_x^2S' - 2S'\partial_xS' - fS' + \partial_x\eta
\end{equation}
where we have defined $S' = \partial_xS$.  We see that $S'$ obeys the same diffusion equation as the particle density in the case of Clifford evolution (see Equation \ref{densevol}), which also served as the derivative of entropy.  We conjecture that this hydrodynamic equation should be valid not just for Clifford-projective dynamics, but for the area-law phase of all models of unitary-projective dynamics in this universality class.

\section{Floquet Random Circuits}
\label{sec:floquet}

We now move from Clifford circuits to a fully random circuit model, similar to Ref. \onlinecite{amos1, amos2}. Specifically, we consider two 1-dimensional $L$-site Floquet (time-periodic) unitary circuits generated by Haar-distributed random unitaries, where the quantum states at each site span a $q$-dimensional Hilbert space. We will be taking a large $q$ limit, corresponding to an infinite-dimensional local Hilbert space.

Our set-up for the unitary-projective time evolution is as follows. The unitary dynamics of the system is modelled by a Floquet circuit. A non-unitary measurement layer is applied after every $p$ layers of the unitary circuit. Each time we apply a measurement layer, we randomly draw $f$ fraction of sites to perform projective measurements. We make use of developments \cite{amos1, amos2}  allowing exact calculation of the ensemble average of exponential of the Renyi-$\alpha$ entropies for $\alpha \geq 2$ with unitary-projective time evolution in the large-$q$ limit using diagrammatic techniques. Remarkably, the diagrammatic approach provides a mapping from ensemble averages of observables to emergent classical statistical mechanics problems.

When working with Floquet random circuits, it is important to be mindful of the order of limits, since there are three separate limits being taken: the thermodynamic limit $L \rightarrow \infty$, the long time limit $t \rightarrow \infty$, and the limit of large onsite Hilbert space dimension $q \rightarrow \infty$. The appropriate order of limits depends of course on the problem we are trying to solve - and the physics is highly sensitive to the order of limits. For example, when the $L\rightarrow \infty$ limit is taken {\it before} the $t \rightarrow \infty$ limit, then one obtains the analysis of ~\onlinecite{SkinnerRuhmanNahum}. In this order of limits, the authors of ~\onlinecite{SkinnerRuhmanNahum} obtain (at least for Renyi-0 entropy) a phase transition between a low measurement volume law phase and a high measurement area law phase. In contrast, the analysis we present is in the opposite order of limits, when $t\rightarrow \infty$ {\it before} $L \rightarrow \infty$. We show that if the limit $q \rightarrow \infty$ is also taken before $L \rightarrow \infty$ then there is an area law saturation of entanglement entropy for any non-vanishing fraction $f$, and no volume law phase. In contrast, if the limit $L \rightarrow \infty$ is taken before the limit of large $q$, then we can only argue for area law saturation at sufficiently large $f$, and cannot exclude the possibility of an area-volume law transition at a critical $f$. 

\subsection{Floquet Haar Random Unitary Circuit}\label{fhruc}
\begin{figure}[t!]
	\centering
	\includegraphics[width=0.475\textwidth]{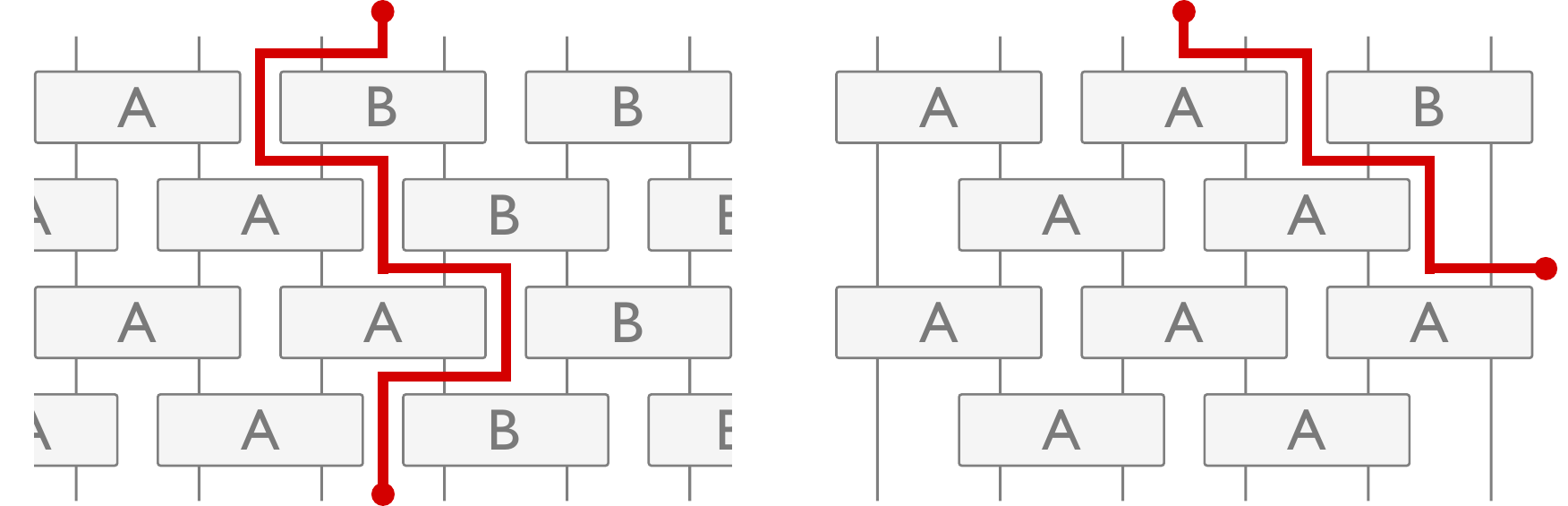}
	\caption{Left: An example of minimal-length DW diagrams for $\langle q^{-S_2(t)} \rangle$ at large $q$  at a time \textit{smaller} than the saturation time $ t_{\text{sat}} = L/2$, which can be generated as follows\cite{amos1}: Draw a DW beginning at the position of bipartition from the top to the bottom; Turn the DW to the left or the right for a distance of lattice spacing  when the DW encounters a 2-gate; Repeat until the DW reaches the bottom of the diagram. In general, there are $2^{t}$ of such diagrams, and each diagram is translated into an algebraic term as $q^{-h} = q^{-t}$ in the $1/q$-perturbative expansion of  $\langle q^{-S_2(t)} \rangle$, where $h$ is the number of horizontal wall segments along the DW. This implies that $S_2(t<L/2) \sim t $.  Right: One of the two minimal-length DW diagrams at a time \textit{larger} than $t_{\text{sat}} $. These two diagrams have DW directed solely to the left or the right, and each contributes a factor of $q^{-L/2}$. This implies that $S_2(t>L/2) \sim L/2 $.}
	\label{fig:dw1}
\end{figure}

First, we review the model and the results of half-system bipartite entanglement spreading without projective measurements. The model is defined by a $q^L\times q^L$ Floquet operator $W=W_2\cdot W_1$, where $W_1 = U_{1,2} \otimes U_{3,4} \otimes  \ldots U_{L-1,L}$ and $W_2 = \mathbf{1}_q \otimes U_{2,3} \otimes U_{4,5} \otimes  \ldots\mathbf{1}_q $. Each $U_{i,i+1}$ is a $q^2 \times q^2$ unitary matrix acting on sites $i$ and $i+1$. In Reference \onlinecite{amos1}, $\langle q^{-S_\alpha(t)} \rangle$ is written as a $1/q$-perturbative series in the large-$q$ limit, and is mapped to an emergent statistical mechanical problem, which, for $\alpha =2$, amounts to generating all minimal-length domain wall (DW) diagrams in Fig.~\ref{fig:dw1}.  The solution gives 
\begin{equation} \label{reneq}
\lim_{q\to \infty }\langle q^{(1 - \alpha) S_\alpha(t)}\rangle  \sim \left\{
\begin{array}{cll}
2^t \,	 q^{(1-\alpha) t} & \quad & t\leq L/2 \\
2  \, q^{(1- \alpha)L/2} & \quad & t >   L /2
\end{array} \right. \; ,
\end{equation}
which suggests a linear growth of $S_\alpha(t<L/2 ) \sim t$ before the saturation time, and a volume-law saturation $S_\alpha(t>L/2 ) \sim L/2$ after the saturation time.

\begin{figure}[t!]
	\centering
	\includegraphics[width=0.475\textwidth]{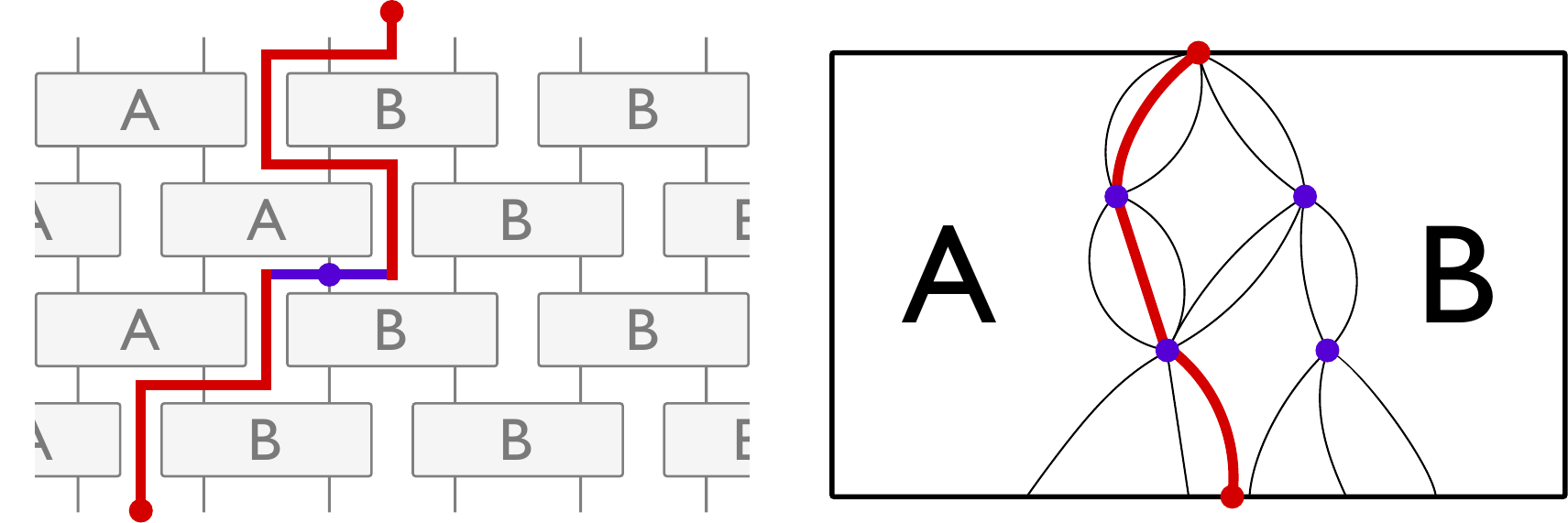}
	\caption{Left: An example of DW diagrams for  $\langle q^{-S_2(t)} \rangle$ at large $q$ and $t \leq t_{\text{sat}}$ with a single measurement at $i$ and time $t_\mathcal{P}$ (represented by the symbols in purple). These diagrams are a subset of the diagrams in Fig.~\ref{fig:dw1} containing those with DW passing through point $i$ and $t_\mathcal{P}$. Each of such diagrams is algebraically translated to $q^{-h + P}$, where $P$ is the number of projection operators along the DW.  Right: A simplified illustration of the set of minimal-length DW diagrams before saturation time with multiple layers of projective measurements.  A leading order diagram (in red) belongs to the set of ``path-integral'' diagrams that scatter off the projective measurements: the DW begins from the top centre of the diagram, walks along one of the minimal-distance paths (in black), and pass through one of the measurements (purple dots) on every measurement layer until it reaches the bottom edge of the diagram.} 
	\label{fig:dw2}
\end{figure}

For the sake of simplicity, we will assume $\alpha =2$ for the following derivations, but the proofs can be straightforwardly extended to general Renyi index $\alpha$.  Now, we investigate the behaviour of entanglement entropy growth with unitary-projective time evolution at a time smaller than the saturation time. It is instructive to consider the effect of performing a single projection operator $\mathcal{P}(c,i) = \sqrt{q} \ket{c,i}\bra{c,i}$ onto color state $c$ at site $i$ and time $t_\mathcal{P}$.  In App.~\ref{proof1}, we prove that the relevant diagrams are the minimal-length DW diagrams whose DW passes through the space-time point $i$ and $t_\mathcal{P}$ (Fig.~\ref{fig:dw2} left), because this segment of DW does \textit{not} give rise to a factor of $q^{-1}$ (i.e. this DW segment is ``free''), which makes this diagram more dominant in the $1/q$-perturbative series.  The algebraic factor associated to such diagrams is $q^{-t + 1}$.  Generally, for a finite period $p$, fraction $f$, and time $t$ smaller than $t_{\text{sat}}$ (to be specified below), the leading diagrams of $\langle q^{- S_2(t)}\rangle$ contain DW that passes through the location of a projection measurement every $p$ unitary layers, and the order is $q^{-t + t//p}$ (Fig.~\ref{fig:dw2} right), where $//$ is the floor division.  An expression for the multiplicity of such leading order DW diagrams can be written in terms of a transfer matrix (acting on a Hilbert space labelled by the DW position) as described in App.~\ref{multi}.

At a time sufficiently large, the minimal-length DW diagrams are the ones with DW ending on the side of the diagrams. Without measurements, there are two leading diagrams where the DW-s are solely directed to the left or the right (Fig~\ref{fig:dw1} right). With measurements, minimal-length DW diagrams are the ones with DW passing through a certain number of projection operators to reach the side of the diagrams.  We prove in App.~\ref{proof1} the following equation.
\begin{align}\label{fruqcres}
\nonumber
\lim_{q\to \infty } &
\langle  q^{ (1- \alpha )S_\alpha(t)} \rangle   
\\
&
=   \left\{ 
\begin{array}{cll}
\beta  \, q^{(1-\alpha )(t - t//p)}  & \quad &  t \leq ps \\
\gamma  \, q^{ (1- \alpha ) [(p-1)(s-1) + t \vspace{-0.1cm} \Mod{p} ]}  & \quad & t > ps
\end{array}
\right. \; ,
\end{align}
where $s =   \text{ceil} (1/2f) $,  ceil$(\cdot)$ is the ceiling function, and where $\beta$ and $\gamma$ are independent of $q$ and dependent of $L$. On the LHS, we have implicitly averaged over the positions of projection operators in a given measurement layer. 

\begin{figure}[t!]
	\centering
	\vspace{2.5mm}
	\includegraphics[width=0.475\textwidth]{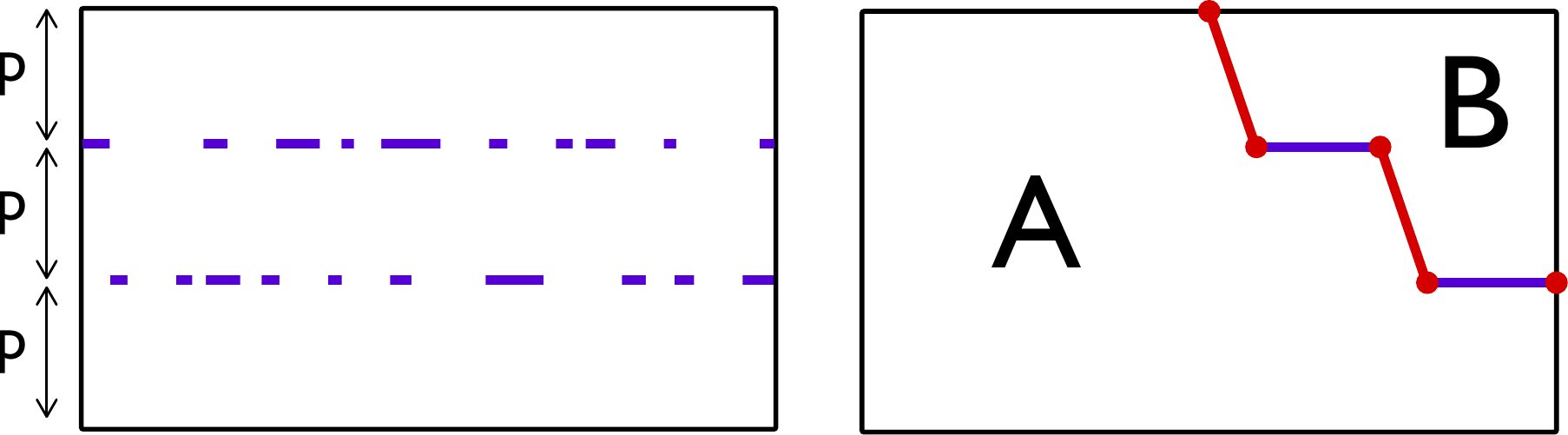}
	\caption{Left: The simplified diagrammatic representation of a realization of projective measurements before ensemble-averaging, where the locations of projective measurements (purple) are scattered randomly along the measurement layers. Right: The minimal-length diagrams are staircase DW diagrams, which requires specific realizations of projection measurements. The DW connects the side of the diagram after $\text{ceil}(1/2f)$ number of periods. Therefore, the order is at least $q^{-(p-1) \text{ceil}(1/2f)}$. The right diagram is drawn for $f=1/4$.}
	\label{fig:staircase}
\end{figure}

The intuition behind the result is can be explained using Fig.~\ref{fig:staircase}. At large $t$, the leading ``staircase'' diagrams are the ones with the DW reaching the side of the diagrams in the shortest distance, utilizing the ``free'' segments of walls provided by the projective measurements (purple segments in Fig.~\ref{fig:staircase}). 
 The area-law saturation originates from the fact that the DW in these leading staircase diagrams pass through $L$ free DW segments.
Take $f=1/4$ as an example, there exists a realization of projection measurements such that the DW can reach the side using 2 measurement layers each of which provides $L/4$ ``free'' DW segments (Fig.~\ref{fig:staircase} right). So the orders of such diagrams are  at least $q^{-2(p-1)}$. In general, it takes $\text{ceil}(1/2f)$ number of ``stairs'' (and hence periods) to reach either side of the diagram. This explains Eq.~\ref{fruqcres}.

The combinatoric factor arising from requiring a staircase configuration of projection measurement locations implies that the coefficient $\gamma$ is suppressed in $L$ as $f^{L}$, but independent in $q$ (to be discussed further in Sec.~\ref{heu}). 
Taking $\log_q$ on both side of Eq.~\ref{fruqcres}, and taking the limit $q \rightarrow \infty$ for fixed but arbitrarily large $L$, we have for large $t > ps$,

\begin{equation} \label{floqarea}
\lim_{q\to \infty } 
S_\alpha  \leq   (p-1)\, \text{ceil}\left(  \frac{1}{2f}  \right) 
\, ,
\end{equation}
which means that the Renyi-$\alpha$ entropy for $\alpha \geq 2$ saturates according to the area law for finite $p$ and non-vanishing $f$ in the limit $q \rightarrow \infty$ at any finite but arbitrarily large $L$.  This is one of the first analytical calculations that demonstrate an area law saturation of entanglement entropy in Floquet random quantum circuit under unitary-projective dynamics, which remarkably occurs even with an infinite local Hilbert space dimension. 

Some questions naturally follow from this analysis.  First, how does the result extend to finite $q$?  And what happens when we take the limit of large  $q$, but {\it after} we take the thermodynamic limit $L \rightarrow \infty$?  To address these questions, we provide a heuristic argument to show that there are exponential many staircases diagrams in Sec.~\ref{heu}, and it is plausible for the area-law saturation to survive at least for large enough $f$, even when we take the limit of large $q$ after we take the thermodynamic limit.

\subsection{Floquet Random Phase Circuit}\label{frpc}
\begin{figure}[t!]
	\centering
	\vspace{2.5mm}
	\includegraphics[width=0.475\textwidth]{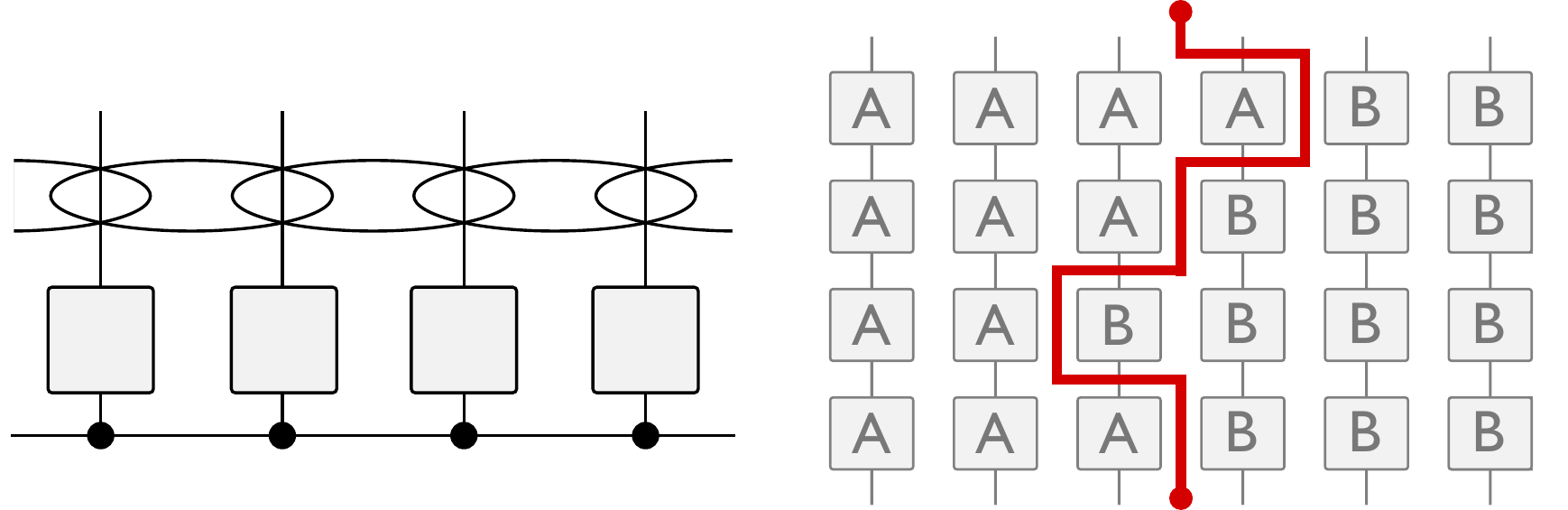}
	\caption{Left: The Floquet random phase circuit where the first layer $W_1$ contains Haar random unitary 1-gates, and the second layer $W_2$ contains diagonal 2-gates with random phases. Right: An example of the leading order minimal-length DW diagrams of $\langle q^{-S_2(t)} \rangle$ in the regime when $\epsilon \gg \log q$. Note that the diagonal 2-gates are not illustrated in this diagram. An analogous description concerning the enumeration and the order of DW diagrams given in Fig.~\ref{fig:dw1} applies here.}
	\label{fig:rp1}
\end{figure}
The domain wall picture extends beyond the Floquet Haar random unitary circuit. Here we describe the analogous result (proved in App.~\ref{app:fprc}) 
in the Floquet random phase circuit first introduced in Ref.~\onlinecite{amos2} (Fig.~\ref{fig:rp1}). This model is similarly defined by a $q^L \times q^L$ Floquet operator $W = W_2 \cdot W_1$ where $W_1 = U_1 \otimes U_2 \otimes \dots \otimes U_L$ is a tensor product of independent Haar random unitaries, and $W_2$ couples neighbouring sites using a diagonal 2-gates with entries
\begin{equation} \label{diagphase}
[W_2]_{a_1, \dots, a_L ; a_1, \dots, a_L} = \exp\left( i \sum_n \varphi_{a_n, a_{n+1}} \right) \; .
\end{equation}
Each $\varphi_{a_n, a_{n+1}}$ is a Gaussianly-distributed random phase with mean zero and variance $\epsilon$. This circuit has two parameters: (i) $q$, which allows us to obtain analytical results at $q \to \infty$, and (ii) $\epsilon$, which allows us to tune how strongly nearest-neighbouring sites couple. 

If the unit time is defined after the application of $W_1$ \textit{and} $W_2$, 
then in the strong-coupling limit $\epsilon \gg \log q$, both Eq.~\ref{reneq} and \ref{fruqcres} apply, and hence we have again Eq.~\ref{floqarea}. This statement is proved in App.~\ref{app:fprc}. 

\subsection{Heuristics: Staircase Diagrams}\label{heu}

\begin{figure}[t!]
	\centering
	\vspace{2.5mm}
	\includegraphics[width=0.475\textwidth]{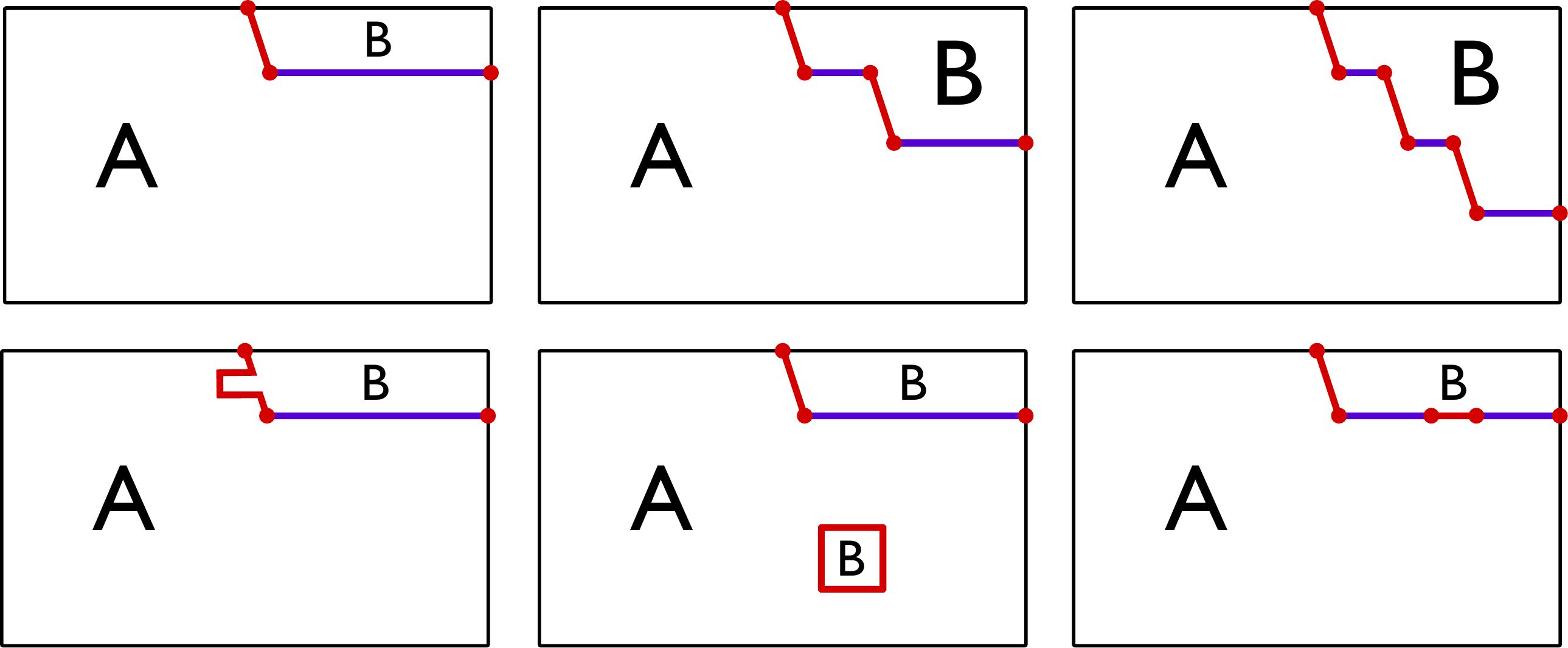}
	\caption{Top: The illustration of $1$-, $2$- and $3$- staircase diagrams for $\langle q^{-S_2(t)} \rangle$ from the left to the right. The orders of the diagrams are $q^{-(p-1) k }$ where $k$ is the number of staircase. The positions of where the staircases begin give rise to the multiplicity of the $k$-staircase diagram, which is proportional to $L^{k-1}/(k-1)!$.  Bottom: Other variants of staircase diagrams which one may expect to contribute to the area law saturation of $S_2$, but for each of these diagrams, one can identify another ``non-Gaussian'' diagram\cite{amos1} that will lead to cancellations. These diagrams is \textit{not} explicitly considered in Eq.~\ref{sumstaircase}.   
	}
	\label{fig:staircasesum}
\end{figure}
We provide a heuristic argument for an exponential number of staircase diagrams, so that it is plausible for the area-law saturation of $S_\alpha$  for $\alpha \geq 2$ to at least survive for large enough $f$, when the limit $L \rightarrow \infty$ is taken before the limit of large $q$.  For the sake of simplicity, we consider an alternative set-up where there is a probability $f$ for each site in a measurement layer to be projectively measured.  Again, we explain the derivation explicitly for $\alpha=2$, but the argument holds for general $\alpha \geq 2$. Lastly, the argument below is applicable to both the models in Sec.~\ref{fhruc} and ~\ref{frpc}.

The origin of area law saturation $S_2$ can be related to DW diagrams that fulfil two criteria: (i) These diagrams have DW starting from the top centre of the diagram and ending on the side of the diagrams (otherwise the order of the diagrams decrease in time for $p> 1$). (ii) These diagrams have DW passing through at least $L$ projective measurements (otherwise the diagram would have an order that scales in $L$). The diagrams that satisfy these criteria are the staircase diagrams (e.g. Fig.~\ref{fig:staircase} right). We call a staircase diagram with $k$ number of staircases a $k$-staircase diagrams. The $1/q$-perturbative series of $\langle  q^{- S_2(t)} \rangle$ can be written in terms of the contribution of the $k$-staircase diagrams (Fig.~\ref{fig:staircasesum} top) as    

\begin{align} 
\label{sumstaircase}
\langle & q^{- S_2(t)} \rangle
\sim 
f^{L} \sum _{k= 1 }^L c_k  \left[ q^{-(p-1)} \right]^k + \dots\\
&\sim  \exp[L ( \log f + q^{-(p-1)})]   \left[ q^{-(p-1)} \right] + \dots
\end{align}

\noindent where $c_k = \int_0^L dl_1 \int_0^{l_1} dl_2 \dots \int_0^{l_{k-2}} dl_{k-1} =L^{k-1}/(k-1)! $ is the multiplicity of the $k$-staircase diagram. As an example, the multiplicity of a 2-staircase diagram is of order $L$ because while the first staircase (counting from the top) always begin in the top center of the digram, the second staircase can begin anywhere between the center and the far right of the diagram (Fig.~\ref{fig:staircasesum} top middle). The dots denote all other contributions to $\langle  q^{- S_2(t)} \rangle$.  For general $\alpha \geq 2$, the above equation becomes
\begin{align} 
\label{sumstaircase2}
\langle  &q^{(1- \alpha) S_\alpha (t)} \rangle \nonumber
\\
&\sim  \exp[L ( \log f + q^{(1- \alpha)(p-1)})]   \left[ q^{(1-\alpha)(p-1)} \right] + \dots
\end{align}

For this contribution to not be suppressed in $L$, we must have the $L$-dependent exponent to be greater than zero,
\begin{equation} \label{fcrit}
f  \gtrsim  e^{- q^{(1- \alpha )(p-1)} }  
\, .
\end{equation}
This contribution implies it is plausible that the area-law saturation of $S_\alpha$ survive at least for large enough $f$.
Note that this argument is not completely rigorous, because we have not systematically looked at all sub-leading terms in the $1/q$-perturbative series of $\langle  q^{(1- \alpha)S_\alpha(t)} \rangle$.  In particular, there can in principle be diagrams that are algebraically translated in negative terms which lead to cancellation with other positive terms (these are called ``non-Gaussian'' diagrams in Ref.~\onlinecite{amos1}). To summarise, we have found an exponential numbers of staircase diagrams and argued that it is plausible for an area-law saturation of $S_\alpha$ to survive at least for large enough $f$, even when the limit $L \rightarrow \infty$ is taken before the limit of large $q$.

\section{Constraints on Volume-Law Phases}
\label{sec:General}

In this section, we present general arguments which highly constrain the form of von Neumann entanglement entropy in systems time-evolving via local unitaries and projectors, assuming a local Hilbert space of finite dimension.  We find that a volume law entanglement entropy is not stable in the presence of measurements unless there is a subleading correction, which should serve as an important signature of measurement physics within the volume law phase.  Insofar as Von Neumann entropy upper bounds Renyi entropies of higher index (e.g. $S_2$), this argument also provides upper bounds on the scaling of higher Renyi entropies.  Throughout this section, `entropy' refers to Von Neumann entropy, unless specified otherwise.

We consider a situation where, in alternating time steps, either nearest-neighbor unitaries or projective measurements are applied to each site with probability f.  The basic observation is that for any region $A$, the rate of entropy increase is only proportional to the size of the boundary $|\partial A|$, while the rate of entropy decrease on $A$ due to measurement can generically be proportional to the total entropy of $A$.  The only way for these rates to balance would be for $A$ to satisfy an area law.  In contrast, a stable volume law phase requires a state in which measurements are much less effective at removing entanglement.  We show that this requirement implies the existence of a subleading term in the entanglement entropy.

\subsection{Constraining strong volume laws}

In this subsection, we exclude the possibility of volume law von Neumann entropy without a subleading correction.  First, we show that any bipartite unitary acting on $D\times D$ dimensions can increase the entanglement entropy between the two parties by no more than $2\log D$.  To see this, imagine such a unitary $U:ab\rightarrow a^\prime b^\prime$ with $d_a=d_{a^\prime}=d_b=d_{b^\prime} = D$ applied to a state $\ket{\psi}_{AaBb}$, where Alice holds systems $Aa$ and Bob holds $Bb$.  Then, the increase in entanglement achieved by applying $U$ to $ab$ is 
\begin{align}
\Delta S_{\rm uni} = &= S(Aa^\prime)_{\rho_{A a^\prime}^f} - S(Aa)_{\rho_{A a^\prime}^i}\\
 & = S(a^\prime|A)_{\rho^f} - S(a|A)_{\rho^i}\\
& \leq S(a^\prime) + S(a) \leq 2 \log D.
\end{align}
Here we have used the notation of conditional entropy $S(A|B) = S(AB)-S(B)$.  We have also used the subadditivity of entropy, $S(AB) \leq S(A) + S(B)$, and the Araki-Lieb inequality: $S(A|B)\geq -S(A)$ (Ref.~ \onlinecite{AL70}).
This is actually a simplified derivation of a bound found in \onlinecite{BHLS03}, which studies the general problem of entanglement generation via bipartite unitaries.

Next, assuming $A = A_1...A_n$ is composed of subsystems, we derive an upper bound on the entropy change caused by measuring a constant fraction $f$ of those subsystems. Letting $T$ be the collection of  subsystems that are \emph{not} measured, and $M_{T^c}$ be the classical outcomes of the measurement on the complement, $T^c$, we see that the average entanglement-entropy change by measurement is given by
\begin{align}\label{Eq:UBDM}
\Delta S_{\rm meas} & = \sum_{T} p_{T} S(A_T|M_{T_c}) - S(A_1...A_n)\\
& \leq  \sum_{T} p_{T} S(A_T) - S(A_1...A_n).
\end{align}

We can use these two observations to conclude that small-scale volume-law-like scaling must saturate to an area law for sufficiently large sizes in any spatial dimension $d$.  For the sake of contradiction, 
 suppose our $A$ consists of $n$ contiguous spins $A_1...A_n$, and that the entropy of the system $A$ scales as
\begin{align}\label{Eq:Scaling}
S(A_1...A_n) & =  \gamma n + g(n)\\
& =  \gamma |A| + g(|A|),
\end{align}
where $g(|A|)= o(|A|)$ is a correction term. 
Our goal is to upper bound 
\begin{align}
\Delta S_{\rm meas} \leq  - \gamma f |A| + o(|A|). \label{eq: measbound}
\end{align}

To this end, we apply Eq.~\ref{Eq:UBDM}, but we must handle a slight subtlety. While Eq.~\ref{Eq:Scaling} posits only the asymptotic behavior of entropies of contiguous sets of spins,  the right-hand side of Eq.~\ref{Eq:UBDM} involves entropies of non-contiguous spins.
To see how this works, fix $T_c$ (the spins being measured) and label the contiguous systems between successive points in $T_c$,  $V_1, ..., V_k$.  The typical size of $V_i$ will be $\approx 1/f$ and there will be $k \approx f n$ such contiguous sets.  

We give two arguments.  In the first argument, we assume a stronger requirement on our correction term, demanding $g(n) = o(1)$.  Given this, fixing a particular $V_i$, letting $A_L$ denote a large number of spins to the left of $V_i$ and $A_R$  a large number to the right, 
strong subadditivity of entropy implies that 
\begin{align}
S(V_i) \leq S(A_LV_i) + S(V_iA_R) - S(A_LV_iA_R) .
\end{align}
Applying the assumed scaling, this becomes
\begin{align}
S(V_i) &\leq \gamma(|A_L| + |V_i|) + \gamma(|V_i| + |A_R| )\\
&    - \gamma(|A_L| + |V_i| + |A_R|) + o(1)\\
&\rightarrow \gamma |V_i|.
\end{align}
As a result, we find that 
\begin{align}
\sum_T p_T S(A_T) &= \sum_T p_T S(V_1 ....V_k) \\
&\leq  \sum_T p_T \sum _i S(V_i)\\
&\leq \sum_T p_T \gamma |V_i|\\
& = (1-f) \gamma n.
\end{align}
Substitution into Eq.\ref{Eq:UBDM} then yields Eq.\ref{eq: measbound} as desired. 

In the second argument, we allow a more relaxed scaling, where we do not require that the correction term $g(n) \sim o(1)$, but only require that the deviations around area law are independent and random (with mean zero) for different
sets and over different realizations of $T$.  In this case we find
\begin{align}
\sum_T p_T S(A_T) &= \sum_T p_T S(V_1 ....V_k) \\
&\leq  \sum_T p_T \sum _i S(V_i)\\
&= \sum_T p_T \sum _i (\gamma |V_i| + g(V_i))\\
& = (1-f) \gamma n + \sum_T p_T \sum _i (g(V_i))\\
& =  (1-f) \gamma n + O(\sqrt{n}).
\end{align}

In both cases, we find 
\begin{align}
\sum_T p_T S(A_T) \leq (1-f) \gamma n + O(\sqrt{n}),
\end{align}
so that the entropy change due to measurement satisfies

\begin{align}
\Delta S_{\rm meas} & = \leq  \sum_{T} p_{T} S(A_T) - S(A_1...A_n)\\
&\leq (1-f) \gamma n - \gamma n + o(n)\\
& = -f \gamma n + o(n).
\end{align}

The change in entanglement entropy caused by one round of local unitaries satisfies
\begin{align}
\Delta S_{\rm uni} & \leq 2 l \log q \leq 2 |\partial A| \log q,
\end{align}
where $l$ is the number of $q\times q$ unitaries that straddle the boundary between $A$ and $B$, which is equal to the length of the boundary of $A$.  Combining this
with the change in the entanglement entropy due to measurement gives us
\begin{align}
\Delta S_{\rm tot} & = \Delta S_{\rm meas} + \Delta S_{\rm uni}\\
 & \leq -f \gamma |A| + 2 |\partial A| \log q + o(|A|).
\end{align}
Note that for sufficiently large $n = |A|$, this becomes negative since $|\partial A|$ scales more slowly than $|A|$ .  As a result, a stable entropy of form Eq.~(\ref{Eq:Scaling}) cannot be achieved.  
In particular, if we hope for volume law scaling of the form $S(A) = \gamma |A|$, we find positive entropy growth rate can only be sustained for 
\begin{align}
2|A|^{\frac{d-1}{d}}\log q \geq \gamma f|A|, 
\end{align}
which requires
\begin{align}
|A| \leq \left(\frac{2\log q}{\gamma f}\right)^{d}.
\end{align}
Alternatively, volume law entanglement must break down around a saturation entropy
\begin{align}
S_{max} \approx \gamma  \left(\frac{2\log q}{\gamma f}\right)^{d}.
\end{align}

We can also show that a strong volume-law behavior is impossible in $1D$ with a simpler argument.  Given a set $A$, the local-unitary steps will tend to increase the entanglement entropy of $A$, while the projective measurements will tend to decrease it.  Our goal is to identify the size at which these competing forces balance out.   To understand the rate of entropy reduction due to measurements, we make some assumptions about the structure of the state on $A$.  In particular, we consider a situation where the entanglement entropy of $A$ is nearly maximal (the state is nearly maximally mixed) and see how large an $A$ is consistent with this.  In a sense, we are asking how big can $A$ be and be consisent with a very strong notion of volume law.  Suppose $A$ has $|A|$ spins.  Then, after one step of measurements, a fraction $f|A|$ spins will be measured, and the resulting entropy will be $(1-f)|A|\log q$, which is an entropy change of $\Delta S_{meas} = -f|A|\log q$.  When a layer of local unitaries is applied, only two of the unitaries will straddle the edges of $A$ (one at each end).  The unitary step will therefore increase the entanglement entropy of $A$ by $\Delta S_{\rm uni} \leq 4 \log q$.  So, after one unitary step and one measurement step, the change in entanglement entropy is
\begin{align}
\Delta S_{\rm tot} & = \Delta S_{\rm meas} + \Delta S_{\rm uni}\\
& \leq 4 \log q -f|A|\log q.
\end{align}
We therefore find $\Delta S_{\rm tot} \leq  0$ for $4 \leq f|A|$.  This suggests that for $|A|\leq 4/f$, unitary-projective dynamics will increase the entanglement entropy of $A$, but that it will saturate around $|A| \approx 4/f$.  This simple argument holds only for near-maximally mixed states on $A$.

\subsection{Logarithmic corrections and phase transitions}
Now we argue that our general argument does allow for logarithmic corrections to area laws, and also for phase transitions between area law phases with and without logarithmic corrections. The argument is simple: suppose the entropy scales as 
\begin{align}
S(A_1...A_n) & =  \gamma |\partial A| \log |n| \label{eq: logscaling}
\end{align}
An argument analogous to that presented in the previous subsection gives
\begin{align}
\Delta S_{meas} &\le \gamma |\partial A| \log (1-f)\\
\Delta S_{tot} & \le 2 |\partial A| \log q -  \gamma |\partial A| \log (1-f)
\end{align}
For $f < f_c = 1 - q^{2/\gamma}$ the upper bound on $\Delta S$ is positive, such that entropy growth of the form Eq.\ref{eq: logscaling} can be sustained indefinitely, leading to a $\log L$ correction to area law behavior. For $f > f_c$, $\Delta S_{tot} < 0$, and the scaling Eq.\ref{eq: logscaling} cannot be sustained, allowing only for a true area law. 


\section{Conclusions}
\label{sec:conclusions}

In this work, we have investigated the entanglement dynamics of a system featuring a combination of unitary and projective time evolution, which have competing effects on quantum entanglement.  We have argued that the effects of projection can keep the system in a state of low entanglement, featuring an area law for entanglement entropy, in contrast with the volume-law entanglement entropy resulting from generic pure unitary time evolution.  We have constructed several toy models which capture the important features of unitary-projective evolution, such as the growth of short-range entanglement due to unitary evolution and the removal of entanglement at any scale by projective measurements.  In the simplest model, described in the language of Bell pairs, an area-law phase persists down to arbitrarily low measurement rates.  We have also shown that, starting from a product state, entanglement can often overshoot its late-time value prior to saturating to the area law.  We then constructed a generalized cluster model which features an area-to-volume law transition at a finite critical measurement rate.  We have tested this intuition in various concrete yet analytically tractable realizations of unitary-projective evolution.  Specifically, we have studied Clifford evolution in one dimensional qubit systems, and Floquet random circuits in one dimension.  In all cases, we find a stable area law phases. In some models, the area law phase persists to arbitrarily weak but non-zero measurement rates, whereas in others it gives way at a critical measurement rate to a low measurement volume law phase. We have further demonstrated that in the low measurement volume law phase there must be a subleading correction to the volume law (i.e. a strong volume law is impossible at any non-zero measurement rate). 

We thus conclude that projective measurements can generically restrict systems to area law entanglement, at least for a sufficiently high measurement rate.  This implies - counter-intuitively - that measurement of a quantum system can {\it inhibit} thermalization through local unitary time evolution, and help keep the system in a low entanglement state. This seems to be rather good news both for Fisher's model of quantum cognition, and for efforts to store and manipulate quantum information more generally.

The results of this manuscript are now consistent with numerical work by Li, Chen, and Fisher \cite{LiChenFisher,LiChenFisher2}, and by Skinner, Ruhman, and Nahum \cite{SkinnerRuhmanNahum}. In Li, Chen, and Fisher, numerical data on unitary-projective evolution in systems of size up to $L=500$ was reported, and a phase transition was observed between a high measurement phase in which entanglement entropy reached a volume law.  Meanwhile, Skinner, Ruhman and Nahum reported numerics on system sizes up to $L=24$, and observed an analogous area-to-volume law transition.

Our work paves the way for future investigations into unitary-projective dynamics.  There is much that remains unknown about the new measurement-driven area-law phase, as well as the area-to-volume law transition.  The transition appears to have an important relationship with certain statistical mechanics models, such as percolation\cite{SkinnerRuhmanNahum}, though the extent to which all properties of the transition can be understood in this language remains unclear.  Regarding the new measurement-driven area-law phase itself, to what extent can the system be understood as an athermal `localized' phase?  Also, since area-law entanglement entropy is more commonly associated with quantum ground states, can the measurement-driven phase host unusual sorts of quantum orders, such as seen in the context of localization protected order?\cite{lpo}  There are many interesting questions remaining to be answered in this exciting new field.

\section*{Acknowledgments}

We acknowledge inspiration for this project from a talk given by Matthew Fisher at the PCTS conference on ``Statistical Mechanics out of Equilibrium" in May 2018, which was funded partially by the Foundational Questions Institute (fqxi.org; grant no. FQXi-RFP-1617) through their fund at the Silicon Valley Community Foundation.  We are also grateful to Yaodong Li for pointing out a crucial error in an earlier version of this manuscript. We acknowledge useful discussions with Yang-Zhi Chou, Mario Collura, Andrea De Luca, Yaodong Li, Xiao Chen, Matthew Fisher, Brian Skinner, Jonathan Ruhman, and Adam Nahum.  This work was supported by NSF Grant 1734006 (GS,MP), by EPSRC Grant No. EP/N01930X/1 (AC), by a Simons Investigator Award to Leo Radzihovsky (MP), by the Foundational Questions Institute (fqxi.org; grant no. FQXi-RFP-1617) through their fund at the Silicon Valley Community Foundation (MP,RN), and by the Alfred P. Sloan foundation through a Sloan Research Fellowship (RN). Some of this work was carried out during the Boulder Summer School for Condensed Matter Physics, which is supported by NSF grant DMR-13001648. R.N. is also grateful to the KITP, which is supported by the National Science Foundation under Grant No. NSF PHY-1748958, and the program `The Dynamics of Quantum Information', where part of this work was completed.

\appendix
\section{Evaluation of $\langle q^{(1-\alpha) S_\alpha(t)}\rangle$ in Section \ref{sec:floquet}}\label{proof1}

\subsection{Floquet Haar Random Unitary Circuit}\label{app:fhruc}

\begin{figure*}[t!]
	\centering
	\includegraphics[angle=0, width=1 \linewidth]{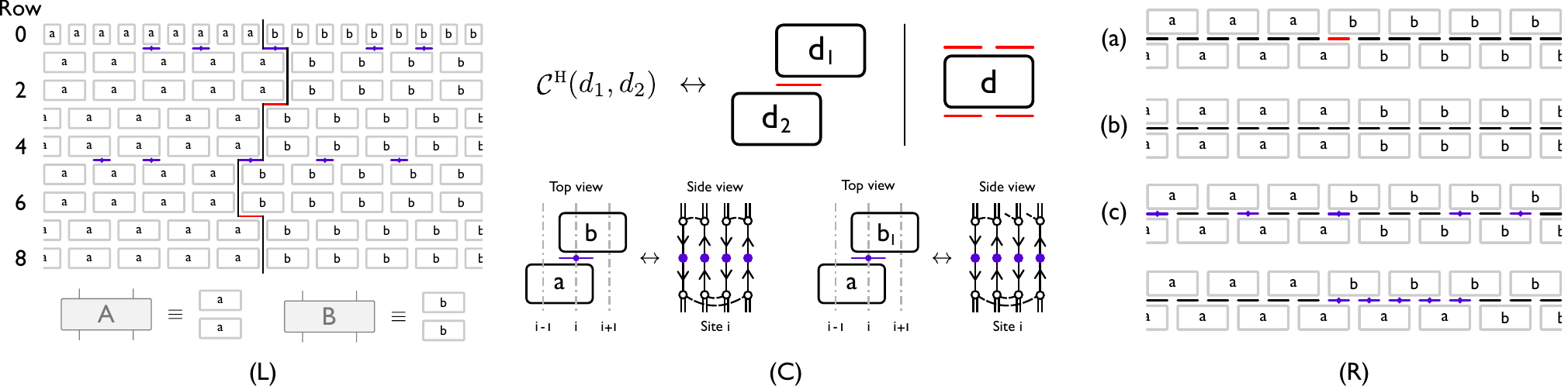}
	\caption{
		(L) top: The block representation of a leading order DW diagram for $\langle q^{-S_2(t)}\rangle $ for the Floquet model specified in Sec.~\ref{fhruc} in the large-$q$ limit in early time $t$. The far left column labels the rows of blocks. The configuration of the top row of blocks is fixed by the boundary condition of the trace structure of the observable $\langle q^{-S_2(t)}\rangle $ (see Ref.~\onlinecite{amos1} for details). 
		We compute $\langle q^{-S_2(t)}\rangle $ by evaluating the partition function of the following ensemble. Each realization in the ensemble has state variables living in each block and local Boltzmann weights between two vertically-neighbouring blocks. Only non-trivial Boltzmann weights (the ones smaller than unity) are drawn in red. The locations of projective measurements are drawn in purple. (L) bottom: A dictionary between the simplified diagrammatic representations in the main text and the ones in the appendices.		
		\\
		(C) top left: The diagrammatic representation of the local Boltzmann weight (in red) between two vertically-neighbouring blocks $d_1$ and $d_2$. 
		(C) top right: A block (not at the edges of a diagram) has four local Boltzmann weight with its neighbouring blocks.		
		(C) bottom: 
		This figure illustrates the derivation of Table~\ref{dwcost}. Ref.~\onlinecite{amos1} is required to understand the figures. (C) bottom left: The LHS is the top view of two vertically-neighbouring block with time axis vertical and space axis horizontal. The RHS is the side view of the block in terms of loops on site $i$ with time axis vertical and space axis pointing out of the page. The purple dots represent the projective operators. 		
		We have $\mathcal{C}^{\text{H}}_{\text{proj}}(a,b ) = 1$, because as long as the top and bottom blocks have local contractions, this region of the diagram has saturated its highest possible order. (C) bottom right: The loop on site $i$ with purple dots is of length at least 3 due to the non-local contraction. This implies that the existence of this loop reduces the overall order of the diagram by $q^{-1}$ from the highest possible order, since the 3-loop could have been split into smaller loops. Consequently, we have $\mathcal{C}^{\text{H}}_{\text{proj}}(a,b_1 ) = q^{-2/3}$.
		\\
		(R): Examples of leading order configurations of different types of row of walls. (a) An even row of walls without projection operators along the rows. (b) An odd row of walls (which cannot have any projection operators). (c) Even rows of walls with projection operators. (c) bottom: If multiple projection operators are located next to each other in space, there can be a leading order configuration in which a DW between domains $a$ and $b$ horizontally extends over a number of sites.
	} \label{frucpanel}
\end{figure*}

In this section, we prove Eq.~\ref{fruqcres} for $\alpha =2$ for the model described in Sec.~\ref{fhruc}.  The case of $\alpha > 2$ is discussed in App.~\ref{app:genalpha}.  We take for granted what is proven in Sec. IV C, App. B 2 and E in Ref.~\onlinecite{amos1}. 
We begin by reviewing the emergent statistical mechanical problem 
described in Sec. IV.C. in Ref.~\onlinecite{amos1} without projective measurements.  

$\langle q^{- S_2(t)}\rangle$ can be expressed as a $1/q$--perturbative series in the large-$q$ limit, which can in turn be mapped to a partition function of the following ensemble at zero temperature.  (This mapping is exact only in the large-$q$ limit.)  The ensemble consists of configurations (diagrammatically represented in Fig.~\ref{frucpanel} L) whose state variables live in blocks and take values from the set $\{a, b,a_1, a_2, b_1, b_2,x \}$ (Fig. 14 in Ref.~\onlinecite{amos1}). Between every pair of vertically-neighbouring blocks $d_1$ and $d_2$, there is a local Boltzmann weight $\mathcal{C}^{\text{H}}(d_1, d_2)$ (explicitly derived and written in Table 1 in Ref.~\onlinecite{amos1}) which is diagrammatically represented as the horizontal boundary (with a width of lattice spacing) between the two blocks (Fig.~\ref{frucpanel} C top). Without projective measurements, the weight is unity if and only if $d_1 = d_2$, so it is useful to distinguish the boundaries between domains of blocks of different values, which we call \textit{domain walls} (DW). 
The associated global Boltzmann weight of diagram $G$ (which we also refer to as the \textit{order} of $G$) is the product of all local Boltzmann weights of walls $w = (d_1, d_2)$ between neighbouring blocks $d_1$ and $d_2$,
\begin{equation} \label{rowcost}
\mathcal{O}(G) = \prod_{\text{walls}} \mathcal{C}^{\text{H}}(w ) \,.
\end{equation}
In the limit $q \rightarrow \infty$, the partition function is dominated by diagrams with the largest Boltzmann weight or the highest order. 
It is proven\cite{amos1} 
that the leading order diagrams are minimal-length DW diagrams with DW separating domains of $A$-blocks and $B$-blocks (Fig.~\ref{fig:dw1}).  

The presence of projection operators effectively provides locations where DW can form without lowering the order of a diagram. In other words, DW-s that pass through projection operators are ``free''. 
To be precise, we state, in Table~\ref{table1}, the local Boltzmann weight function $\mathcal{C}^{\text{H}}_{\text{proj}}(\cdot, \cdot)$ for two vertically-neighbouring blocks that sandwich a projection measurement in-between. This function is derived using the same method introduced in Ref.~\onlinecite{amos1}
and two examples 
are provided in Fig.~\ref{frucpanel} C bottom. Importantly, $\mathcal{C}^{\text{H}}_{\text{proj}} (\cdot, \cdot)$ differs $\mathcal{C}^{\text{H}} (\cdot, \cdot)$ in the following way: aside from the diagonal entries of the table, there is a single entry, namely $(a,b)$, in the table of $\mathcal{C}^{\text{H}}_{\text{proj}}(\cdot, \cdot)$ that gives a Boltzmann weight of unity. 
This implies that the projection operators effectively provide locations at which DW can form without reducing the overall order of the diagram. 
We will use this observation to show that $S_2(t)$ saturates to an area law in late time in the large-$q$ limit. 

We proceed in the proof with three steps: 
(i) We analyse the diagrams row-by-row, and show an upper bound in the order for each row of walls (which is defined by two neighbouring rows of blocks). (ii) We identify the leading order diagrams by invoking the ``sink-source'' arguments\cite{amos1}, and by showing that these diagrams saturate the bounds found in (i). (iii) We show that all diagrams with the highest order are algebraically translated into positive factors (so there can be no cancellation between these contributions).

We label each row of blocks as in the far left of Fig.~\ref{frucpanel} L, and a row of walls 
by the label of the row of blocks above. For step (i), consider 3 types of rows of walls: (a) Even rows of walls without projection operators along the rows (e.g. row 2 in Fig.~\ref{frucpanel} L); (b) Odd rows of walls which cannot have any projection operators (row 3 in Fig.~\ref{frucpanel} L); (c) Even rows of walls with projection operators (row 4 in Fig.~\ref{frucpanel} L). 

The following upper bounds in the order of rows of type (a) and (b) are proved in Ref.~\onlinecite{amos1} using Table 1 in the reference.
For case (a), 
if there are two types of blocks, $a$ and $b$, on the top row of \textit{blocks}, the upper bound of the order of the row of \textit{walls} is $q^{-1}$, given rise by a single factor of $\mathcal{C}^{\text{H}}(a,b)$ (while all the other local Boltzmann weights are $\mathcal{C}^{\text{H}}(a,a) = \mathcal{C}^{\text{H}}(b,b) = 1$, see Fig.~\ref{frucpanel} R (a)). Note that if there is only a single type of blocks, say $b$, along the top row of blocks, then the upper bound of unity is always saturated by choosing the bottom row of blocks identical to the top one, i.e. also $b$.
For case (b), regardless of the number of block types in the top row of blocks, one can always find a configuration of row of walls with order unity, by choosing the bottom row of blocks identical to the top row of blocks (Fig.~\ref{frucpanel} R (b)).
For case (c), the upper bound of order is unity even if there are two types of blocks, say $a$ and $b$, on the top row of blocks (c.f. case (a)), because the DW between domains of block $a$ and $b$ can occur at the position of the projective measurement. Furthermore, depending on the realization of positions of projection operators, a leading row of walls can be an extended segments of horizontal DW (Fig.~\ref{frucpanel} R (c) bottom). This concludes step (i).

\begin{table}[t!]
	\begin{tabular}{ l | l | l | l | l | l | l | l} \label{table1}
		$\omega$ &  $a$ & $b$ & $a_1$ & $a_2$ & $b_1$ & $b_2$ & $x$\\
		\hline
		%
		$a$     &  1 & 1 & $q^{-1/2}$ & $q^{-1/2}$ & $q^{-2/3}$ & $q^{-2/3}$ & $q^{-1}$\\
		$b$     &               & 1 & $q^{-2/3}$ & $q^{-2/3}$ & $q^{-1/2}$ & $q^{-1/2}$ & $q^{-1}$\\
		$a_1$ &               &              & 1 & $q^{-1}$ & $q^{-3/4}$ & $q^{-3/4}$ & $q^{-1/2}$\\
		$a_2$ &               &              &              & 1 & $q^{-3/4}$ & $q^{-3/4}$ & $q^{-1/2}$\\
		$b_1$ &               &              &              &              & $1$ & $q^{-1}$ & $q^{-1/2}$\\
		$b_2$ &               &              &              &              &              & 1 & $q^{-1/2}$\\
		$x$     &               &              &              &              &              &              &1\\
	\end{tabular}
	\caption{Upper bounds for the local Boltzmann weight $\mathcal{C}^{\text{H}}_{\text{proj}}(\cdot, \cdot)$ associated with the boundaries between two vertically-neighbouring blocks that sandwich a projective measurement in-between. The matrix is symmetric and so only the upper triangle is written explicitly. The upper bounds are saturated by all Boltzmann weights that appear in the leading order diagrams of $\langle q^{-S_2(t)}\rangle $ in the large-$q$ limit.
		Note in particular that $\mathcal{C}^{\text{H}}_{\text{proj}} (a,b)= 1$, while $\mathcal{C}^{\text{H}}(a,b) = q^{-1}$ in Ref.~\onlinecite{amos1} \label{dwcost}}
\end{table}

To find candidates of leading order diagrams , we invoke the ``sink-source'' argument introduced in Ref.~\onlinecite{amos1}: 
Suppose we assign an orientation to a wall (e.g. if there are only domains $a$ and $b$ in the diagram, we can choose a DW to be directed forward if domain $a$ is on its left and $b$ on its right.) A \textit{source} is a point in the diagram from which a outwardly-directed DW has to originate. For example, the center top of the Fig.~\ref{frucpanel} L has a source, because regardless of whether the block immediately below is of type $a$ or $b$, a DW has to be generated. 
A \textit{sink} is similarly defined. Importantly, a DW originated from a source must end at a sink. Due to this argument, for $\langle q^{-S_2(t)}\rangle $, there must be a DW coming from the center top of Fig.~\ref{frucpanel} L and ending either (1) along the bottom edge of the diagram (Fig.~\ref{fig:dw2} right), or (2) on the side of the diagram (Fig.~\ref{fig:staircase} right). Since the order of a diagram decreases as the DW length increases, the \textit{minimal-length} DW diagrams of types (1) and (2) are candidates for leading order diagrams.

Now we identify the highest order diagrams of types (1) and (2). For type (1), there exists a minimal-length DW diagram (as illustrated in Fig.~\ref{fig:dw2} and Fig.~\ref{frucpanel} L) that saturates the highest order $q^{-1}$ on every row of walls:
Every even row of walls without measurements saturates the highest order $q^{-1}$ (as in (a) in Fig.~\ref{frucpanel} R), and every odd row and even row with measurement saturate the highest order of unity (as in (b) and (c) in Fig.~\ref{frucpanel} R). Therefore, the highest order diagram of type (1) has an order $q^{-t + t//p}$, where $t//p$ is the number of measurement layer the DW passes through. 


For type (2), recall that we are averaging over a separate ensemble of measurements over their positions in Eq.~\ref{fruqcres} (the other average is over the Haar ensemble). 
In this average, 
there are realizations of the circuit that have projective measurements forming a stair-case configuration as in Fig.~\ref{fig:staircase} right (the purple lines). These realizations are suppressed in $L$ as $f^{L}$ but nevertheless are the dominant contributions in the limit $q \rightarrow \infty$ for fixed but arbitrarily large $L$ (the limit $L \rightarrow \infty$ is discussed in Sec.~\ref{heu}).
The minimal-length DW for such configuration reaches the side of the diagram with at most $\text{ceil}(1/2f)$ numbers of period. 
Diagrams of this type are leading order diagrams because each of them saturates the highest order bound for each row that form the staircase (in a similar way to the case of type (1) above). To find the order of the leading diagrams, we count the number of even rows of wall without projective operators, and obtain the exponent in the second case of Eq.~\ref{fruqcres}.

At early time, type (1) diagrams provide the leading order diagrams because type (2) diagrams do not exist due to insufficient number of staircases. 
At sufficiently late time, type (2) diagrams dominates because the order of diagrams of type (2) does not scale in $t$. In Ref.~\onlinecite{amos1}, it is proven that the \textit{only} leading order diagrams are of types (1) and (2) in early and late time respectively (the appearance of projective measurements only trivially change the proof of this statement in Ref.~\onlinecite{amos1}). The time of the regime-change between (1) and (2) is determined by 
the time when $s=\text{ceil}(1/2f)$ number of staircases can form. This gives $t^* = p s$.  This concludes step (ii).

Finally, we check (iii) to ensure the leading diagrams do not translate into algebraic terms that cancel each other out.  To this end, note that the leading order diagrams of type (1) and (2) always have odd rows of walls of type (b) in Fig.~\ref{frucpanel} R. Such diagrams are called ``Gaussian''\cite{amos1}, and are algebraically translated into positive contributions to $\langle q^{- S_2(t)}\rangle$.  We have therefore proved Eq.~\ref{fruqcres}.

\subsection{Floquet Random Phase Circuit}\label{app:fprc}

\begin{figure*}[t!]
	\centering
	\includegraphics[angle=0, width=1 \linewidth]{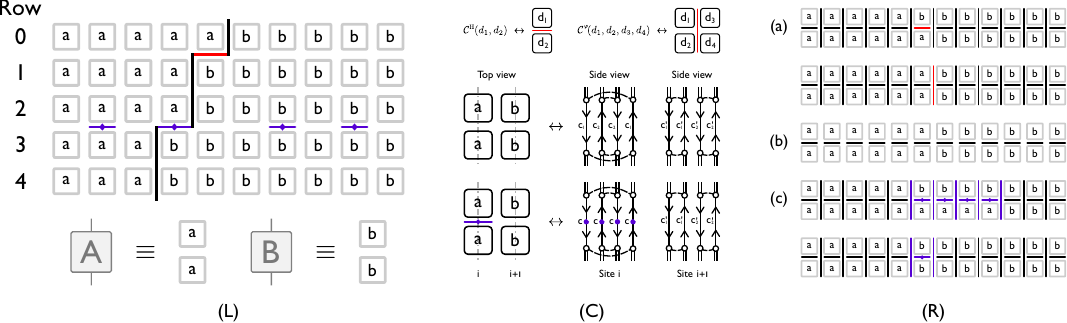}
	\caption{
		(L) top: The block representation of a leading order DW diagram for $\langle q^{-S_2(t)}\rangle $ for the Floquet model specified in Sec.~\ref{frpc} in the large-$q$ limit in early time $t$. The convention is identical to the one specified in Fig.~\ref{frucpanel} L. (L) bottom: A dictionary between the simplified diagrammatic representations in the main text and the ones in the appendices.		
		\\
		(C) top left: The average over Haar-random unitaries give rise to the local Boltzmann weight $\mathcal{C}^{\text{H}  }(d_1, d_2)$  between vertically-neighbouring blocks $d_1$ and $d_2$, which is diagrammatically represented as a horizontal line in red. 
		(C) top right: The average over the Gaussianly-distributed phases gives rise to an  Interaction-Round-a-Face type Boltzmann weight $\mathcal{C}^{ \varphi  }(d_1, d_2,d_3,d_4 )$, which is diagrammatically represented as a vertical line in red.
		(C) middle and bottom: 
		These figures illustrate two example derivations of $\mathcal{C}^{ \varphi  }(\cdot, \cdot, \cdot, \cdot)$ and $\mathcal{C}^{ \varphi  }_{\text{proj}}(\cdot, \cdot, \cdot, \cdot)$. The convention is the same as the one given in Fig.~\ref{frucpanel}. Ref.~\onlinecite{amos1} and \onlinecite{amos3} are required to understand the figures. 
		(C) middle: Without measurements, the corresponding phases of this quadruplet are $\exp [i (\varphi_{c_1,c'_1} - \varphi_{c_2,c'_1} + \varphi_{c_1,c'_2} - \varphi_{c_2,c'_2} )]$. Since these phases do not explicitly cancel each other, we have  $\mathcal{C}^{\varphi}(a,a,b,b) = e^{-2\epsilon}$ in the large-$q$ limit.
		(C) bottom: In the presence of projective measurements (purple), We have $\mathcal{C}^{\varphi}_{\text{proj}}(a,a,b,b) = 1$, because 
		the phases exactly cancel each other out. 
		\\
		(R): Examples of configurations of different types of rows of walls. (a) Even rows of walls without projection operators along the rows. 
		(a) top: If the top row of blocks has two types of blocks, $a$ and $b$, then the only order unity rows of walls are the ones where the DW walk one lattice spacing to the left or the right.  
		(a) bottom: A configuration where all factors of $\mathcal{C}^{\text{H}}$-s are unity, but there is a factor of $\mathcal{C}^{ \varphi  }(a, a, b, b) = e^{-2 \epsilon }$ which make the configuration sub-leading.
		(b) A leading order odd row of walls (which cannot have any projection operators). (c) Two leading order even rows of walls with projection operators. (c) bottom: If multiple projection operators are located next to each other in space, there can be a leading order configuration in which a DW between domains $a$ and $b$ horizontally extends over a number of sites.
	} \label{frpcpanel}
\end{figure*}
The corresponding proof of Eq.~\ref{fruqcres} for $\alpha =2$ for  the Floquet random phase circuit  specified in Sec.~\ref{frpc} and Fig.~\ref{fig:rp1} is very similar to the one in App.~\ref{app:fhruc}, except that in this model there are two types of local Boltzmann weights to account for.  (The case of $\alpha > 2$ is discussed in App.~\ref{app:genalpha}.)  First we review the evaluation of $\langle q^{- S_2(t)}\rangle$  in the absence of projective measurements.  In the large-$q$ limit, $\langle q^{- S_2(t)}\rangle$ can be mapped to the partition function of an ensemble whose realizations have the state variables living in blocks (Fig.~\ref{frpcpanel} L). There are two types of local Boltzmann weights: The average over Haar-random unitaries gives rise to $\mathcal{C}^{\text{H}}(d_1, d_2 )$ between every pair of vertically-neighbouring blocks $d_1$ and $d_2$ (Fig.~\ref{frpcpanel} C top left and Table 1 in Ref.~\onlinecite{amos1}). The average over the Gaussianly-distributed phases gives rise to an  Interaction-Round-a-Face type Boltzmann weight $\mathcal{C}^{ \varphi  }(d_1, d_2,d_3,d_4 )$ (Fig.~\ref{frpcpanel} C top right and Ref.~\onlinecite{amos3}). 
The associated global Boltzmann weight or order of diagram $G$ is the product of all local Boltzmann weights. In the large-$q$ limit and in the strong-coupling regime where $\epsilon \gg \log q$, it can be proven\cite{amos3} that the leading diagrams are DW diagrams where the DW walks a unit lattice spacing to the left or to the right below every layer of Haar-random unitaries as in Fig.~\ref{fig:rp1} right. If we choose the convention where a unit time is defined after the application of $W_1$ and $W_2$, we recover Eq.~\ref{reneq}.
 
Now we describe the derivation of $\mathcal{C}^{\varphi}(\cdot, \cdot, \cdot, \cdot)$ without  measurements, and $\mathcal{C}_{\text{proj}}^{\varphi}(\cdot, \cdot, \cdot, \cdot)$ in the presence measurements. 
Consider a quadruplet of blocks $(d_1,d_2,d_3,d_4)$. Due to the coupling term described in Eq.~\ref{diagphase}, there are four random phases encoded in this quadruplet (see an example below). In Ref.~\onlinecite{amos3}, the Boltzmann weight $\mathcal{C}^{\varphi}(\cdot, \cdot, \cdot, \cdot)$ is derived based on the following observation: For each random phase $\exp(i \varphi_{c_1,c'_1})$ that is not explicitly cancelled by another phase with the same labels, a factor of $e^{-\epsilon /2}$ arises from the integral over the random phases in the large-$q$ limit. As an example, consider $(a,a,b,b)$ in Fig.~\ref{frpcpanel} C middle, the associated four random phases can be written as $\exp [i (\varphi_{c_1,c'_1} - \varphi_{c_2,c'_1} + \varphi_{c_1,c'_2} - \varphi_{c_2,c'_2} )]$. Since these phases do not explicitly cancel each other, we have  $\mathcal{C}^{\varphi}(a,a,b,b) = e^{-2\epsilon}$ in the large-$q$ limit.

Suppose there is a projective measurement, say, on site $i$, within the region represented by $(d_1,d_2,d_3,d_4)$. The measurement  projects the states on site $i$ to be in the same state $c$, and consequently provide a new mechanism for phase cancellation among the four phases encoded in the quadruplet. Consider again the example of $(a,a,b,b)$ as in Fig.~\ref{frpcpanel} C bottom, due to the projection onto state $\bar{c}$ on site $i$, all phases are cancelled out and therefore, the associated Boltzmann weight is unity (in contrast to $e^{-2\epsilon }$ without the projective measurements). $\mathcal{C}^{\varphi}(\cdot, \cdot, \cdot, \cdot)$ and  $\mathcal{C}_{\text{proj}}^{\varphi}(\cdot, \cdot, \cdot, \cdot)$ can be derived by looking at a finite number of possible combinations of quadruplets.

We proceed with the proof with steps (i-iii) specified in App.~\ref{app:fhruc}. 
For step (i), we identify the upper bounds in order for the three types of rows of walls in the strong coupling regime $\epsilon \gg \log q$. For rows of walls of type (a), if there are two types of blocks, say $a$ and $b$, on the top row of blocks, it is shown in Ref.~\onlinecite{amos3} that the upper bound is $q^{-1}$, and the only rows that saturate this bound are given in Fig.~\ref{frpcpanel} R (a) top. For type (b), as in App.~\ref{app:fhruc}, the leading order is unity and it is saturated only by rows of walls that are sandwiched between two identical rows of blocks. For type (c), the upper bound in order is unity, since the projection measurement provide a site at which both types of local Boltzmann weights are 1. 
Two examples that saturate this bound are given in Fig.~\ref{frpcpanel} R (c). In particular, as before, there are leading order rows of walls in which a DW between domain $a$ and $b$ extends horizontal over multiple sites.  This concludes step (i). 

The derivation of steps (ii) and (iii) are identical to the ones given in App.~\ref{app:fhruc}. This concludes the proof.

\subsection{Generalization to Higher Renyi Entropies}\label{app:genalpha}
For $\alpha >2$, while it is difficult to compute the multiplicity of leading order diagrams for $\langle q^{- S_2(t)}\rangle$ before saturation time, the order of the leading order diagrams (which is our main interest) are known\cite{amos1}. The proofs for $\alpha \geq 2$ can be straightforwardly extended from the proof for $\alpha =2$ as follows: In step (i) of App.~\ref{app:fhruc} and \ref{app:fprc}, the upper bound for rows of type (a) for general $\alpha$ is $q^{(1-\alpha)}$ instead of $q^{-1}$. In step (ii), the leading diagram candidates remain the same, except that they saturate the new upper bound in order on every odd rows of walls without projective measurements. Step (iii) is identical, and therefore, Eq.~\ref{fruqcres} follows.

\section{Multiplicity of Diagrams for $\langle q^{- S_2(t)}\rangle$ for Small $t$ in Section \ref{sec:floquet}}\label{multi}
In this section, we use a transfer matrix to write an expression for the multiplicity of diagrams of $\langle q^{- S_2(t)}\rangle$  at $t \leq ps$ for a \textit{fixed} realization of the  positions of projective measurements.  The vertical segments of a DW live on the bonds between neighbouring sites. If we label the bond between site $x$ and $x+1$ as the $x$-th bond (for open boundary condition, we label the bond on the left of site 1 as 0, and the one on the right of site $L$ as $L$), then a basis for the Hilbert space of DW is $\ket{x}$, where $x= 0,1, \dots, L$.  In the absence of projective measurements, the multiplicity of all possible minimal-length diagrams of  $\langle q^{- S_2(t)}\rangle$ can be generated by a transfer matrix that maps $\ket{x}$ to $\ket{x-1}$ and $\ket{x+1}$ with a weight of unity at each time step. In the presence of measurements, a projection operator at site $i$ can map $\ket{i-1}$ to $\ket{i}$, and  $\ket{i}$ to $\ket{i-1}$. If there is only a single projective measurement $\mathcal{P}(i)$ at site $i$ and time $t_\mathcal{P}$, we can write the multiplicity as
\begin{equation}
\langle q^{- S_2(t)}\rangle =  \sum_{x_f = 0}^{L} \bra{x_f} T^{t - t_\mathcal{P}} \mathcal{P}(i) T^{t_\mathcal{P}} \ket{L/2} q^{-t + 1}
\end{equation}
where $T$ is the $L+1$ by $L+1$ transfer matrix given by
\begin{equation}
T
=	
\begin{bmatrix}
0    & 1   &         &         \\
1    & 0   & 1       &              \\
& 1   & 0   &  \cdots        \\
&    &   \cdots  &   \cdots       \\        
\end{bmatrix},
\;\;
\mathcal{P}(i) = \ket{i}\bra{i-1} + \ket{i-1}\bra{i} 
\, .
\end{equation}
This approach is generalizable to a diagram with multiple projective measurements. However, complication arises when there are multiple projective measurements at neighbouring sites on the same measurement layer. For instance, if there are measurements at both sites $i$ and $i+1$, then there will be additional terms like $\ket{i+1}\bra{i-1}$, which shifts the DW by two lattice spacings.

\end{document}